\documentclass[aps,pre,reprint,superscriptaddress]{revtex4-1}

\usepackage{amsmath,amsfonts,amsmath,mathtools,bbm,bm}
\usepackage{graphicx}
\usepackage{braket}
\usepackage[colorlinks,linkcolor=blue,urlcolor=blue,citecolor=blue]{hyperref}
\usepackage[all]{hypcap}
\usepackage{dsfont}
\setcitestyle{numbers,square}

\newcommand{\Tr}{\mathrm{Tr}}

\def\a{\alpha}
\def\be{\beta}
\def\g{\gamma}
\def\de{\delta}

\def\Re{\text{Re}}
\def\({\left (}
\def\){\right )}

\newcommand*{\defeq}{\mathrel{\vcenter{\baselineskip0.5ex\lineskiplimit0pt\hbox{\scriptsize.}\hbox{\scriptsize.}}}=}

\begin{document}

\title{Out-of-time-order correlations and the fine structure of eigenstate thermalisation}

\author{Marlon Brenes}
\email{brenesnm@tcd.ie}
\affiliation{Department of Physics, Trinity College Dublin, Dublin 2, Ireland}

\author{Silvia Pappalardi}
\email{silvia.pappalardi@phys.ens.fr}
\affiliation{Laboratoire de Physique de l’\'Ecole Normale Supérieure, 75005 Paris, France}

\author{Mark T. Mitchison}
\affiliation{Department of Physics, Trinity College Dublin, Dublin 2, Ireland}

\author{John Goold}
\affiliation{Department of Physics, Trinity College Dublin, Dublin 2, Ireland}

\author{Alessandro Silva}
\affiliation{SISSA, Via Bonomea 265, I-34135 Trieste, Italy}
\date{\today}

\begin{abstract}
Out-of-time-order correlators (OTOCs) have become established as a tool to characterise quantum information dynamics and thermalisation in interacting quantum many-body systems. It was recently argued that the expected exponential growth of the OTOC is connected to the existence of correlations beyond those encoded in the standard Eigenstate Thermalisation Hypothesis (ETH). We show explicitly, by an extensive numerical analysis of the statistics of operator matrix elements in conjunction with a detailed study of OTOC dynamics,  that the OTOC is indeed a precise tool to explore the fine details of the ETH.  In particular, while short-time dynamics is dominated by correlations, the long-time saturation behaviour gives clear indications of an operator-dependent energy scale $\omega_{\textrm{GOE}}$ associated to the emergence of an effective Gaussian random matrix theory. We provide an estimation of the finite-size scaling of $\omega_{\textrm{GOE}}$ for the general class of observables composed of sums of local operators in the infinite-temperature regime and found linear behaviour for the models considered.
\end{abstract}

\maketitle

\section{Introduction}
\label{sec:intro}

Understanding how isolated quantum systems thermalise under unitary evolution is a theme as old as quantum mechanics itself~\cite{Schrodinger:1927,Vonneumann:1929}. This line of study has seen significant renewed interest in the past decades~\cite{Rigol:2008,Polkovnikov:2011,eisert2015quantum,Borgonovi:2016,Alessio:2016} primarily due to advances in ultra-cold atom physics~\cite{Kinoshita:2006,Lewenstein:2007,Bloch:2012} which have allowed the observation of coherent dynamics over long times. 

In isolated classical systems, thermalisation relies on the emergence of chaos and ergodicity, which together lead phase-space trajectories starting from the same energy to become indistinguishable when averaged over time~\cite{lebowitz1973modern}. The equivalent notion of indistinguishability in quantum many-body systems is provided by the Eigenstate Thermalisation Hypothesis (ETH)~\cite{Deutsch:1991,Srednicki:1994,Alessio:2016}, which states that nearby energy eigenstates cannot be distinguished by local observations. More precisely, the ETH requires the matrix elements of few-body observables in the eigenbasis of a many-body Hamiltonian to obey the following ansatz, which implies thermalisation
\cite{Srednicki:1999, Alessio:2016}:
\begin{align}
\label{eq:eth}
O_{n m} = O(\bar{E}) \delta_{n m} + e^{-S(\bar{E}) / 2}f_{\hat O}(\bar{E}, \omega)R_{n m}\ .
\end{align}
Both $O(\bar{E})$ and $f_{\hat O}(\bar{E},\omega)$ are smooth functions of their arguments, $\bar{E}=(E_n+E_m)/2$, $\omega=E_m-E_n$, $S(\bar{E})$ is the microcanonical entropy and 
$R_{nm}$ are the matrix elements of a random \it statistical matrix \rm with zero average and unit variance. 

More recently, thermalisation has been explored from a new quantum information perspective, with emphasis on the notion of scrambling~\cite{swingle2018}. 
Information scrambling is a more primordial feature of quantum dynamical systems where information, initially stored locally, gets dynamically distributed in global degrees of freedom \cite{Hosur2016Chaos}. This process is explained as a consequence of the growth of operator complexity under time evolution~\cite{Parker19}. Although traditional tools can hardly be of any help in studying this phenomenon, a variety of ideas have emerged recently for this task. Among them, the out-of-time-order correlator (OTOC)~\cite{larkin1969quasiclassical}, suggested to characterise synthetic analogues of black-holes~\cite{swingle2018,shenker2014black,maldacena2016}, has arisen as an important figure of merit for scrambling, ergodicity and quantum chaos in complex many-body quantum systems. Several experimental studies with a variety of platforms have demonstrated that OTOCs indeed characterise scrambling following the operation of a unitary circuit~\cite{garttner2017,li2020,Nie20,mi2021}. 

Recently, Foini and Kurchan~\cite{Foini2019} argued that correlations between the matrix elements of operators in the energy eigenbasis must exist in the ETH [Eq.~\eqref{eq:eth}] to account for the positive exponential growth rate of OTOCs in chaotic models~\cite{maldacena2016}. Based on this result, Murthy and Srednicki~\cite{Murthy19} were able to derive known bounds on the growth rate from the ETH. Chan et al.~\cite{Chan2019} showed that in locally interacting systems the butterfly effect for OTOCs implies a universal form for these correlations. The existence of frequency-dependent correlations has recently been confirmed by Richter et al.~\cite{Richter2020} and a distinction with a regime in which these correlations vanish was identified, by a numerical investigation of the statistical distributions of matrix elements in non-integrable systems.

It remains an open question to establish if these frequency-dependent correlations can be observed in the dynamics of OTOCs and if the timescales associated with late-time chaos can be connected to the presence, or lack thereof, of matrix-element correlations. It is still not clear if temperature plays a role and, furthermore, the scaling as a function of system size of the frequency scale that divides correlated and uncorrelated regimes has yet to be estimated.

In this work, we carry out a thorough study of the frequency and energy dependence of the statistics of off-diagonal matrix elements and of the OTOCs of extensive observables in two experimentally relevant models: hardcore bosons with dipolar interactions in a harmonic trap~\cite{Khatami2013} and an Ising chain with longitudinal and transverse fields~\cite{Kim2013}. In all instances, the statistical matrix appears to have some common features. The matrix elements $R_{nm}$ at a given energy $\bar{E}$ and frequency $\omega$ obey Gaussian statistics~\cite{Beugeling2015, Leblond:2019, Khaymovich2019, BrenesLocal2020, Richter2020, Leblond:2020, Santos2020, BrenesFreq2020}, in contrast with non-ergodic systems \cite{BrenesFreq2020, Luitz2016Anomalous, Luitz2016Long, Foini2019Eigenstate}. We demonstrate that this feature persists well-beyond the infinite-temperature limit. We also further characterise the statistical correlations between $R_{nm}$ at well-separated frequencies. However, these correlations disappear between matrix elements close to the diagonal, indicating the emergence of random-matrix-like behaviour at small frequencies $|E_n-E_m|<\omega_{\rm GOE}$, where $\omega_{\rm GOE}$ is a model- and operator-dependent energy scale, as first demonstrated for non-extensive observables in Ref.~\cite{Richter2020}. We show explicitly that this rich structure is naturally reflected in the dynamics of OTOCs. A comparison between the OTOCs computed on a thermal ensemble and those computed assuming the ETH with a random uncorrelated Gaussian statistical matrix shows convergence of the two on time scales that appear to be related to $\omega_{\rm GOE}^{-1}$. We use this observation to provide an estimation of the scaling as a function of the system size of $\omega_{\rm GOE}^{-1}$ in the infinite-temperature regime. This suggests that an experimental study of OTOCs could be an efficient way to probe the energy scales beyond which a complex, interacting system displays Gaussian random-matrix behaviour in local observables.   

\section{Models and observables}

To address the generic behaviour of thermalising systems that is independent of microscopic details, we consider two different non-integrable models: the first describing hard-core bosons with dipolar interactions in a harmonic trap~\cite{Khatami2013}, while the second is a quantum Ising chain with both transverse and longitudinal fields~\cite{Kim2013}. The Hamiltonian of the first model is ($\hbar \defeq 1$)
\begin{align}
\label{eq:h_hb}
\hat{H}_{\textrm{HB}} = -J\sum_{i=1}^{L-1} \left( \hat{b}^{\dagger}_i \hat{b}^{\phantom{\dagger}}_{i + 1} + \textrm{H.c.} \right) + \sum_{i < l} \frac{V \hat{n}_{i} \hat{n}_{l}}{|i - l|^3} + \sum_i g x_i^2 \hat{n}_i
\end{align}
for a one-dimensional chain with $L$ sites where $\hat{b}^{\dagger}_i$ and $\hat{b}^{\phantom{\dagger}}_i$ are hard-core bosonic creation and annihilation operators, respectively, at site $i$, $\hat{n}_i = \hat{b}^{\dagger}_i \hat{b}^{\phantom{\dagger}}_i$ and $x_i=|i-L/2|$. Hereafter, all energies are given in units of the hopping amplitude $J$ and we set the strength of the dipolar interaction and confining potential to be $V = 2$ and $g = 16 / (L - 1)^2$, respectively (parameters selected from Ref.~\cite{Khatami2013}). The system conserves the total number of bosons, which is guaranteed from $[\hat{H}_{\textrm{HB}}, \sum_i\hat{n}_i]=0$. This symmetry is resolved throughout this work. We focus on the half-filled sub-sector, in which the Hilbert space dimension is $\mathcal{D} = L!/[(L/2)!(L/2)!]$. To avoid parity (spatial inversion) or reflection (spin inversion) symmetries, we add a small perturbation  $\delta\hat{n}_1$ to the Hamiltonian ($\delta = 0.1J$). 

The second model has the following Hamiltonian:
\begin{align}
\label{eq:h_is}
\hat{H}_{\textrm{IS}} = \sum_{i=1}^{L} w \hat{\sigma}^x_i + \sum_{i=1}^{L} h \hat{\sigma}^z_i + \sum_{i = 1}^{L - 1} J \hat{\sigma}^z_{i} \hat{\sigma}^z_{i+1}\ .
\end{align}
We measure energies in units $J$ and set $w = \sqrt{5} / 2$, $h = (\sqrt{5} + 5) / 8$ (see Ref.~\cite{Kim2013}). The only known symmetry associated to this model is parity. We consider the even parity sub-sector for chains with an even number of sites, with a corresponding Hilbert space dimension $\mathcal{D} = 2^L - [(2^L - 2^{L/2})/2]$. 

We consider extensive observables, composed of sums of local operators spanning the entire system
\begin{align}
\label{eq:obs_ext}
\hat{B}_{\textrm{HB}} = \frac{1}{L}\sum_i [ 1 + (-1)^i ]\hat{n}_{i} \ , \quad 
\hat{B}_{\textrm{IS}} = \frac{1}{L}\sum_i \hat{\sigma}^z_{i}\ .
\end{align}
A detailed study of the diagonal matrix elements and two-point correlation functions of these operators demonstrates the validity of the ETH in both models (see Appendix~\ref{ap:diag} for further details).

It is crucial to recognise that, within the ETH, the one- and two-point correlation functions in time do not depend on the details of the statistical matrix $R_{nm}$. In particular, two-point correlators are determined by the smooth function $f_{\hat{O}}(\bar{E},\omega)$ entering Eq.~\eqref{eq:eth}, which itself depends on the variance of matrix elements $O_{nm}$ near a given energy $\bar{E}$ and frequency $\omega$~\cite{Khatami2013,Beugeling2015, Mondaini2017,BrenesFreq2020,BrenesLocal2020, Leblond:2020, BrenesFisher2020}. We refer the reader to Appendix \ref{ap:twopoint}, where we observe excellent agreement between the dynamics of two-point functions computed with respect to a single eigenstate and the canonical ensemble at the same average energy, without any particular considerations about the statistical matrix $R_{nm}$, other than its mean and variance.

The precise distribution of these elements, as well as correlations between matrix elements at different frequencies, thus encode the fine structure of the ETH beyond linear-response theory~\cite{Foini2019}. In the following, we investigate how this structure influences the dynamics of higher-order correlators such as the OTOC.

\begin{figure}[t]
\fontsize{13}{10}\selectfont 
\centering
\includegraphics[width=1\columnwidth]{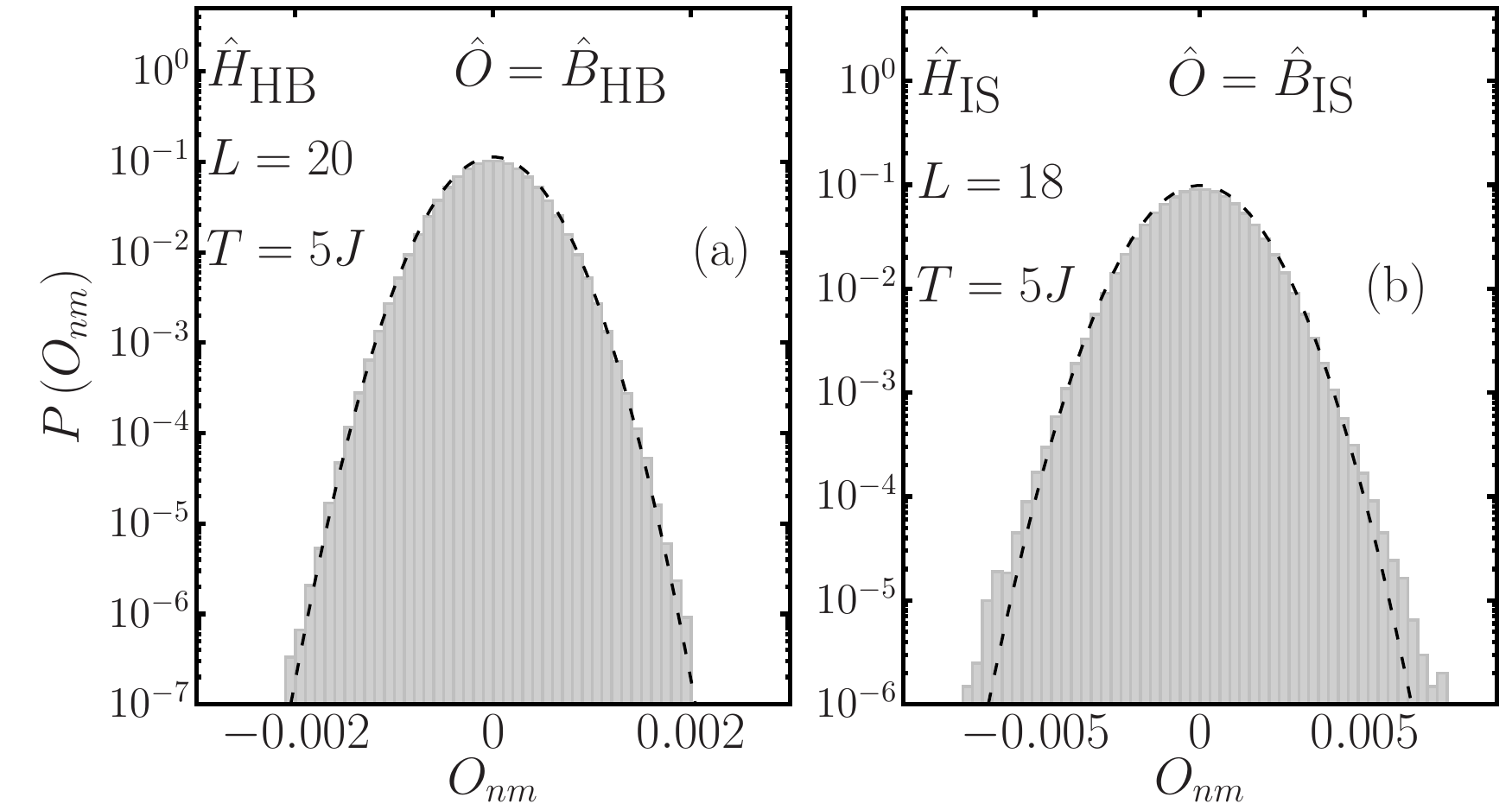}
\caption{Probability distributions of off-diagonal matrix elements in a small frequency range $\omega \lesssim 0.05$. The average energy $\bar{E}$ selected is consistent with a finite canonical temperature $T = 5J$. The distributions are shown in (a) for $\hat{B}_{\textrm{HB}}$ and in (b) for $\hat{B}_{\textrm{IS}}$. Results obtained for finite-sized systems of $L = 20$ for $\hat{H}_{\textrm{HB}}$ and $L = 18$ for $\hat{H}_{\textrm{IS}}$. Dashed lines depict a Gaussian distribution with the same mean and variance.}
\label{fig:m1}
\end{figure}

\section{Gaussian statistics}

The first step towards a statistical characterisation of $\hat{O}$ is to understand the distribution of its individual matrix elements.
Since the number of matrix elements is very large even at small system sizes, we begin by studying the distribution of off-diagonal matrix elements $O_{nm} = \braket{E_n | \hat{O} | E_m}$ in a small frequency-resolved window $\omega \lesssim 0.05$ and a finite temperature $T = \beta^{-1} = 5J$ ($k_B\defeq 1$). The temperature is calculated by associating the average energy $\bar{E}$ with a canonical density matrix $\hat{\rho} = e^{-\beta\hat{H}}/Z$ as $\bar{E} = \textrm{Tr}[\hat{\rho} \hat{H}]$, with $Z = \textrm{Tr}[e^{-\beta\hat{H}}]$. The probability distribution can then be inferred by creating a histogram of all the matrix elements that satisfy $\bar{E} = \textrm{Tr}[\hat{\rho} \hat{H}]$ and $\omega < 0.05$.
In Fig.~\ref{fig:m1} we show the probability distributions obtained by this procedure. The matrix elements are Gaussian-distributed for the extensive operators in both of the models we have studied, as has previously been found for other models and observables in the infinite temperature regime~\cite{Beugeling2015, Leblond:2019, Khaymovich2019, BrenesLocal2020, Leblond:2020, Santos2020, BrenesFreq2020}.

In order to understand if this property pertains to the entire spectrum away from zero frequency and if the same distributions are observed at all temperatures where the ETH is expected to hold, we proceed to evaluate the frequency-dependent ratio~\cite{Leblond:2019}
\begin{align}
\label{eq:gamma}
\Gamma_{\hat{O}}(\omega) \defeq \overline{|O_{nm}|^2} / \overline{|O_{nm}|}^2,
\end{align}
where the averages are performed over a small frequency window $\delta\omega = 0.05$. Should the individual matrix elements be Gaussian-distributed with zero mean at a given value of $\omega$, then $\Gamma_{\hat{O}}(\omega) = \pi / 2$. 
For this particular analysis, we consider $\omega = E_m - E_n$ over the entire spectrum, while the average energy $\bar{E} = (E_n + E_m) / 2$ is chosen to be compatible with a corresponding canonical temperature. The quantity is computed over small bins in $\omega$ of a given size and within a small energy window of width $0.05\epsilon$, where $\epsilon \defeq E_{\textrm{max}} - E_{\textrm{min}}$ is the bandwidth of the Hamiltonian. The average over the small energy window is carried out to account for finite-size eigenstate-to-eigenstate fluctuations.  

\begin{figure}
\fontsize{13}{10}\selectfont 
\centering
\includegraphics[width=1\columnwidth]{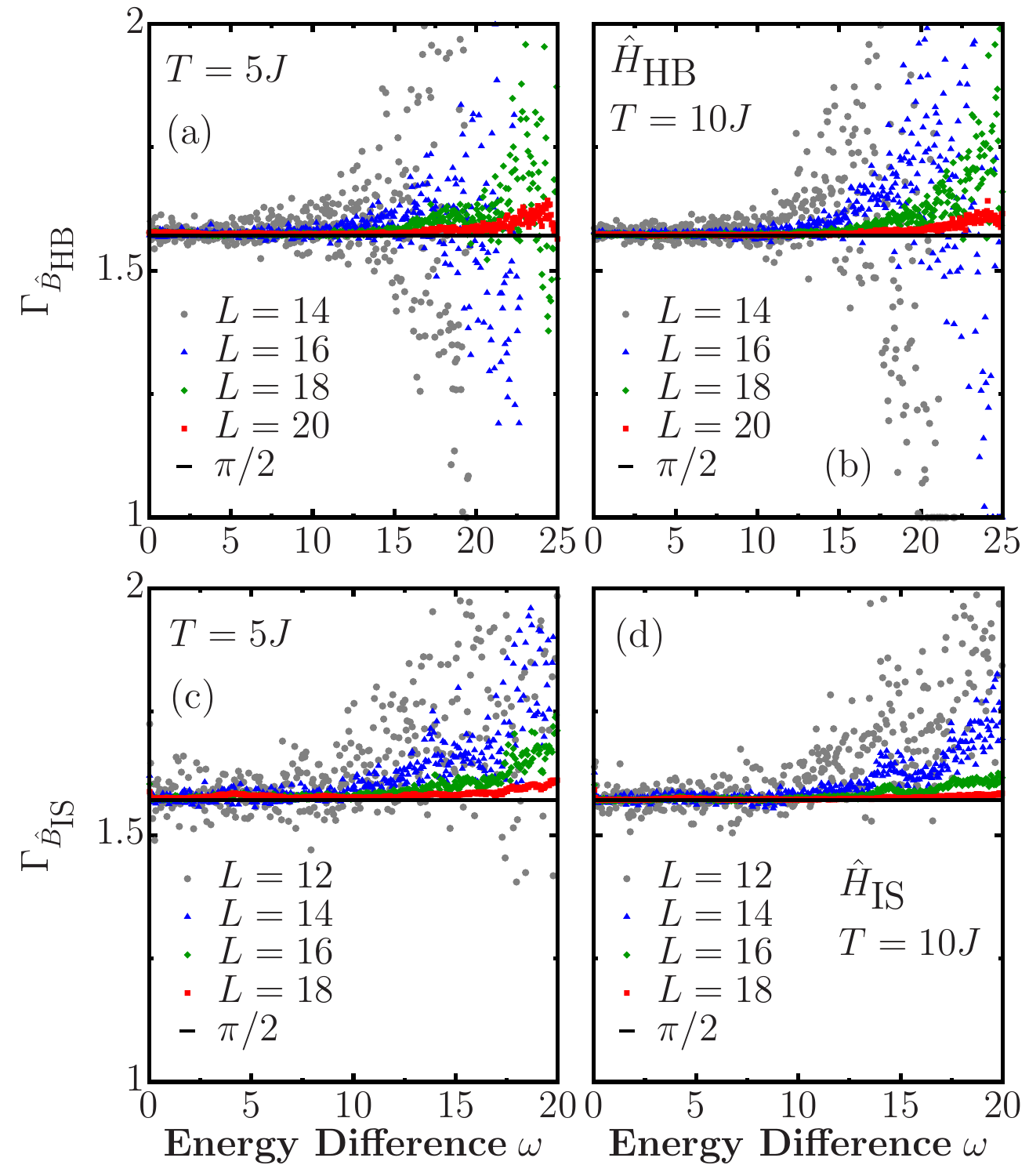}
\caption{$\Gamma_{\hat{O}}(\omega)$, from Eq.~\eqref{eq:gamma}, for operators $\hat{B}_{\textrm{HB}}$ [(a) and (b)] and $\hat{B}_{\textrm{IS}}$ [(c) and (d)] in the eigenbasis of $\hat{H}_{\textrm{HB}}$ and $\hat{H}_{\textrm{IS}}$, respectively. Two different finite temperatures were chosen, $T = 5J$ [(a) and (c)] and $T = 10J$ [(b) and (d)]. The black horizontal line shows the value $\Gamma_{\hat{O}}(\omega) = \pi / 2$. The matrix elements were computed in a small energy window $0.05\epsilon$ where $\epsilon \defeq E_{\textrm{max}} - E_{\textrm{min}}$, and a frequency window $\delta \omega = 0.05$.}
\label{fig:m2}
\end{figure}

In Fig.~\ref{fig:m2} we show the $\Gamma_{\hat{O}}(\omega)$ ratio as a function of $\omega$ and of the system size $L$ for both $\hat{H}_{\textrm{HB}}$ [panels (a) and (b)] and $\hat{H}_{\textrm{IS}}$ [panels (c) and (d)], evaluated for the operators $\hat{B}_{\textrm{HB}}$ and $\hat{B}_{\textrm{IS}}$ from Eq.~\eqref{eq:obs_ext} and for two different temperatures $T = 5J$ and $T = 10J$. We have chosen to display our results for values of temperature away from the infinite-temperature regime, although we have confirmed similar results for other temperature values ($T = 3J$ and $T = 1000J$). 
Gaussian statistics emerge at all frequencies, i.e.~ $\Gamma_{\hat{O}} \approx \pi / 2$ for increasing values of $\omega$ as the system size increases. These findings, together with recent results that have highlighted normality in the distributions of off-diagonal matrix elements in the high-temperature limit~\cite{Leblond:2019,Leblond:2020,BrenesLocal2020,BrenesFreq2020}, strongly suggest that Gaussianity is ubiquitous in non-integrable models for which the ETH is expected to hold, even at finite temperature.

\section{Matrix-element correlations}

Let us now examine the overall structure of the statistical matrix $R_{nm}$ as a function of the mean energy $\bar{E}$ and frequency $\omega$. In particular, we are interested in correlations between matrix elements at different frequencies, which are encoded in the eigenvalue distribution of the matrix $O_{nm}$. In the absence of correlations, the eigenvalue distribution should coincide with that of the Gaussian orthogonal ensemble (GOE), where each matrix element is an independent, identically distributed random variable~\cite{mehta2004random}. Therefore, any deviation from the GOE prediction heralds the presence of correlations between matrix elements.

\begin{figure}[b]
\fontsize{13}{10}\selectfont 
\centering
\includegraphics[width=1\columnwidth]{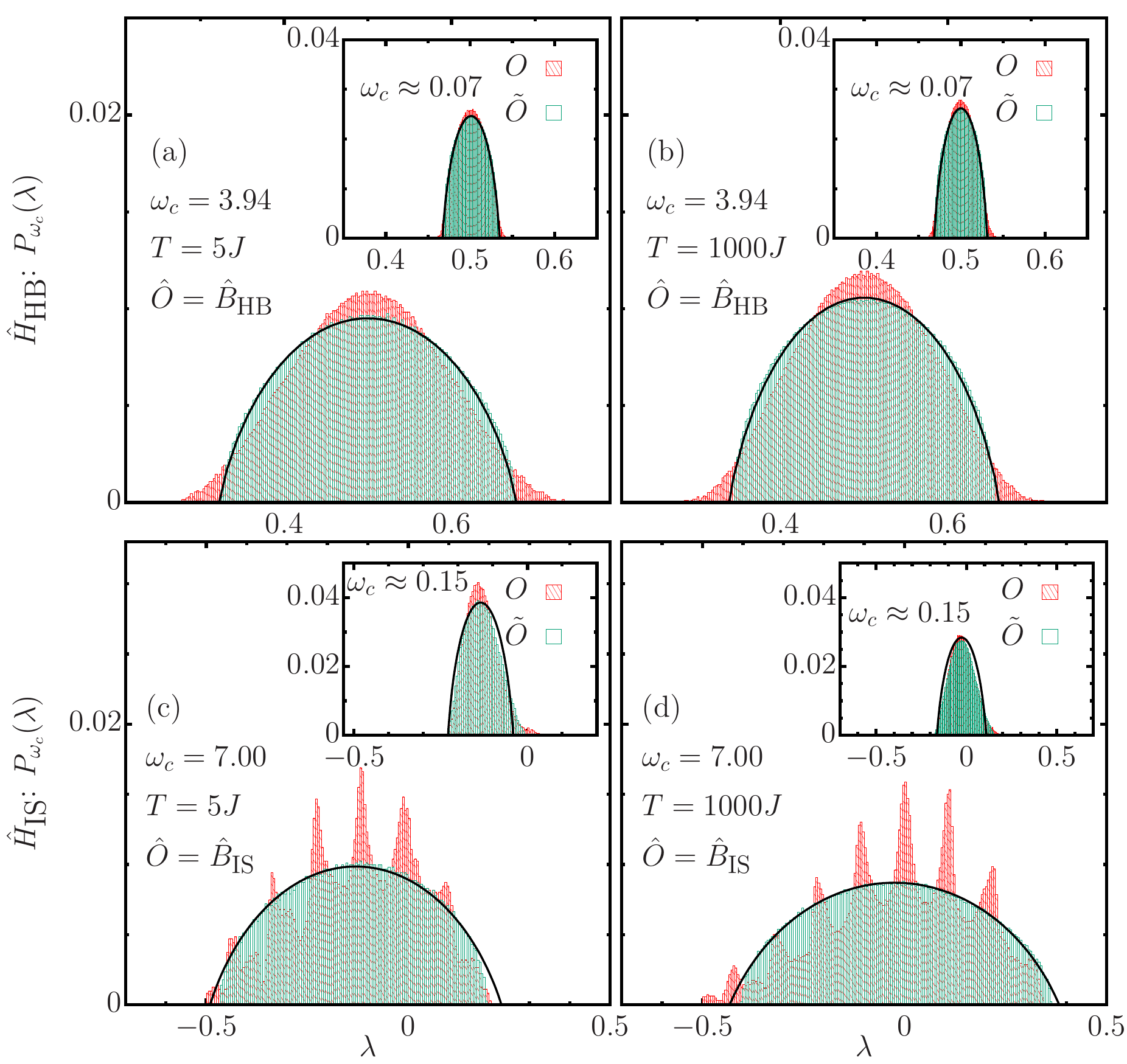}
\caption{Probability distributions $P_{\omega_c}(\lambda)$ of the eigenvalues of the full and randomised sub-matrices [Eqs.~\eqref{eq:o_wc} and~\eqref{eq:o_tilde}] in an energy window $0.125\epsilon$ for $\hat{B}_{\textrm{HB}}$ [(a), (b)] and $\hat{B}_{\textrm{IS}}$ [(e), (f)]. In [(c), (d)] and [(g), (h)] we show $P_{\omega_c}(\lambda)$ for the banded operators $\hat{B}_{\textrm{HB}}$ and $\hat{B}_{\textrm{IS}}$, respectively, with the smallest cutoff frequency $\omega_c$ such that the eigenvalue distribution follows the GOE (see Appendix~\ref{ap:mean}). All the panels on the left correspond to $T = 5$, while $T = 1000$ for the panels on the right. We show the results for $L = 18$ in $\hat{H}_{\textrm{HB}}$ and $L = 16$ in $\hat{H}_{\textrm{IS}}$.}
\label{fig:m3}
\end{figure}

In order to investigate the temperature- and frequency-dependence of such correlations, we consider sub-matrices of $\hat{O}$ restricted to a finite frequency window and construct the corresponding eigenvalue distributions, following Ref.~\cite{Richter2020}. To fix the temperature, we first extract a $\mathcal{D}^{\prime} \times \mathcal{D}^{\prime}$ sub-matrix from $\hat{O}$, centred around the diagonal matrix element $O_{nn}$ such that $E_n = \textrm{Tr}[\hat{\rho} \hat{H}]$. The size $\mathcal{D}'$ of this sub-matrix is selected to encompass an energy range of width $0.125\epsilon$, where $\epsilon \defeq E_{\textrm{max}} - E_{\textrm{min}}$ is the bandwidth. We then further restrict our attention to frequencies $|\omega| < \omega_c$ by setting
\begin{align}
\label{eq:o_wc}
O^{\omega_c}_{nm} \defeq \begin{cases} O_{nm}, \, &\textrm{if}\;  |E_m - E_n| < \omega_c \\ 0, &\textrm{otherwise.} \end{cases}
\end{align}
To test for correlations between these matrix elements, we follow the procedure introduced in Ref.~\cite{Richter2020}: we generate a sign-randomised matrix from the original sub-matrix
\begin{align}
\label{eq:o_tilde}
\widetilde{O}^{\omega_c}_{nm} \defeq \begin{cases} O^{\omega_c}_{nm}, \, &\textrm{probability} = 1/2, \\ -O^{\omega_c}_{nm}, \, &\textrm{probability} = 1/2, \end{cases}
\end{align}
where we apply the sign randomisation on the elements $n \neq m$ to retain the mean and support of the original sub-matrix.
The random sign destroys correlations between matrix elements, leading to the semi-elliptical eigenvalue distribution that is characteristic of the GOE~\cite{mehta2004random,Livan2018introduction}. Comparing the eigenvalue distributions of $O^{\omega_c}_{nm}$ and $\widetilde{O}^{\omega_c}_{nm}$ thus probes correlations between the matrix elements of $\hat{O}$ within a frequency range controlled by the cutoff $\omega_c$. 

The distribution of all the $\mathcal{D}^{\prime}$ eigenvalues $\lambda^{\omega_c}_{\alpha}$ of $\hat{O}^{\omega_c}$ is expressed as
\begin{align}
P_{\omega_c}(\lambda) = \frac{1}{\mathcal{D}^{\prime}} \sum_{\alpha=1}^{\mathcal{D}^{\prime}} \delta \left( \lambda - \lambda^{\omega_c}_{\alpha} \right),
\end{align}
where all the individual $\delta(\cdot)$ peaks are collected in small bins to describe a given probability distribution. The function $P_{\omega_c}(\lambda)$ can be studied as a function of $\omega_c$ and yields a semi-circular distribution if the eigenvalues are uncorrelated. If correlations are to arise, deviations from a semi-circle distribution are observed.

The eigenvalues $\{\lambda\}$ of the sub-matrices in Eq.~\eqref{eq:o_wc} and Eq.~\eqref{eq:o_tilde} are evaluated numerically and the corresponding distributions, $P_{\omega_c}(\lambda)$, are shown in Fig.~\ref{fig:m3} for extensive operators. The eigenvalues of the entire sub-matrix within the chosen energy window show a departure from the semi-elliptical distribution (Fig.~\ref{fig:m3}[(a),(b)] for $\hat{B}_{\textrm{HB}}$ and Fig.~\ref{fig:m3}[(c),(d)] for $\hat{B}_{\textrm{IS}}$), signalling substantial correlations between matrix elements at significantly different frequencies. These correlations are seen for high ($T = 1000J$) [Fig.~\ref{fig:m3}(b,d)] and low  ($T = 5J$) [Fig.~\ref{fig:m3}(a,c)] temperatures alike. For smaller values of the cutoff $\omega_c$, however, the eigenvalue distributions begin to resemble the GOE prediction (Fig.~\ref{fig:m3}[(a),(b)] insets for $\hat{B}_{\textrm{HB}}$ and Fig.~\ref{fig:m3}[(c),(d)] insets for $\hat{B}_{\textrm{IS}}$). Our data are therefore consistent with a crossover to Gaussian random-matrix-like behaviour at low frequencies~\cite{Richter2020}. The frequency scale of the crossover can be estimated from the value of $\omega_c$ at which the distributions appear to coincide with the GOE prediction, $\omega_c = \omega_{\rm GOE}$. Note that, for even smaller values of $\omega_c$, the eigenvalue statistics eventually become Poissonian due to well-known localisation effects~\cite{Alessio:2016}. The insets in Fig.~\ref{fig:m3} display the eigenvalue distributions for smallest frequency values which are still above the localised regime (see Appendix~\ref{ap:mean} for details). This indicates that $\omega_{\textrm{GOE}}$ refers to a different frequency scale, as first denoted in Ref.~\cite{Richter2020}.

\section{OTOC dynamics}

We now study the implications of the results from the previous section for the observable dynamics. As discussed above, two-point correlation functions are independent of the statistical correlations between matrix elements. It is thus crucial to examine higher-order correlators and the OTOC is a natural example. We focus in particular on the squared commutator
\begin{align}
\label{eq_sc}
c(t) \defeq - \left ( \langle [ \hat O(t), \hat O]^2 \rangle - \langle [ \hat O(t), \hat O]\rangle^2 \right ).
\end{align}
To detect the dynamical effect of matrix-element correlations, we compute $c(t)$ in two different ways: {\it i)}~by a thermal average in the canonical ensemble at temperature $T$, and {\it ii)}~using a single eigenstate $\ket{E_n}$ and assuming independent, identically distributed (IID) Gaussian statistics for $R_{nm}$ in the ETH Eq.~\eqref{eq:eth}~\cite{Cotler2017Chaos, Foini2019, Murthy19}, which leads to an expression for the OTOC in terms of two-point functions. Under this approximation, we have 
\begin{align}\label{ETH_Uncorrelated}
[c(t)]_{\textrm{ETH Unc.}} \approx 2 |F_2(0)|^2 - 2 |F_2(t)|^2,
\end{align}
where 
\begin{align}
F_2(t) \defeq \langle \hat{O}(t) \hat{O}(0) \rangle_c \defeq \langle \hat{O}(t) \hat{O}(0) \rangle - \langle \hat{O}(t) \rangle \langle \hat{O}(0) \rangle.
\end{align}
We refer the reader to Appendix~\ref{app:otoc} for details on the derivation leading to this approximation. 

\begin{figure}[t]
\fontsize{13}{10}\selectfont 
\centering
\includegraphics[width=1\columnwidth]{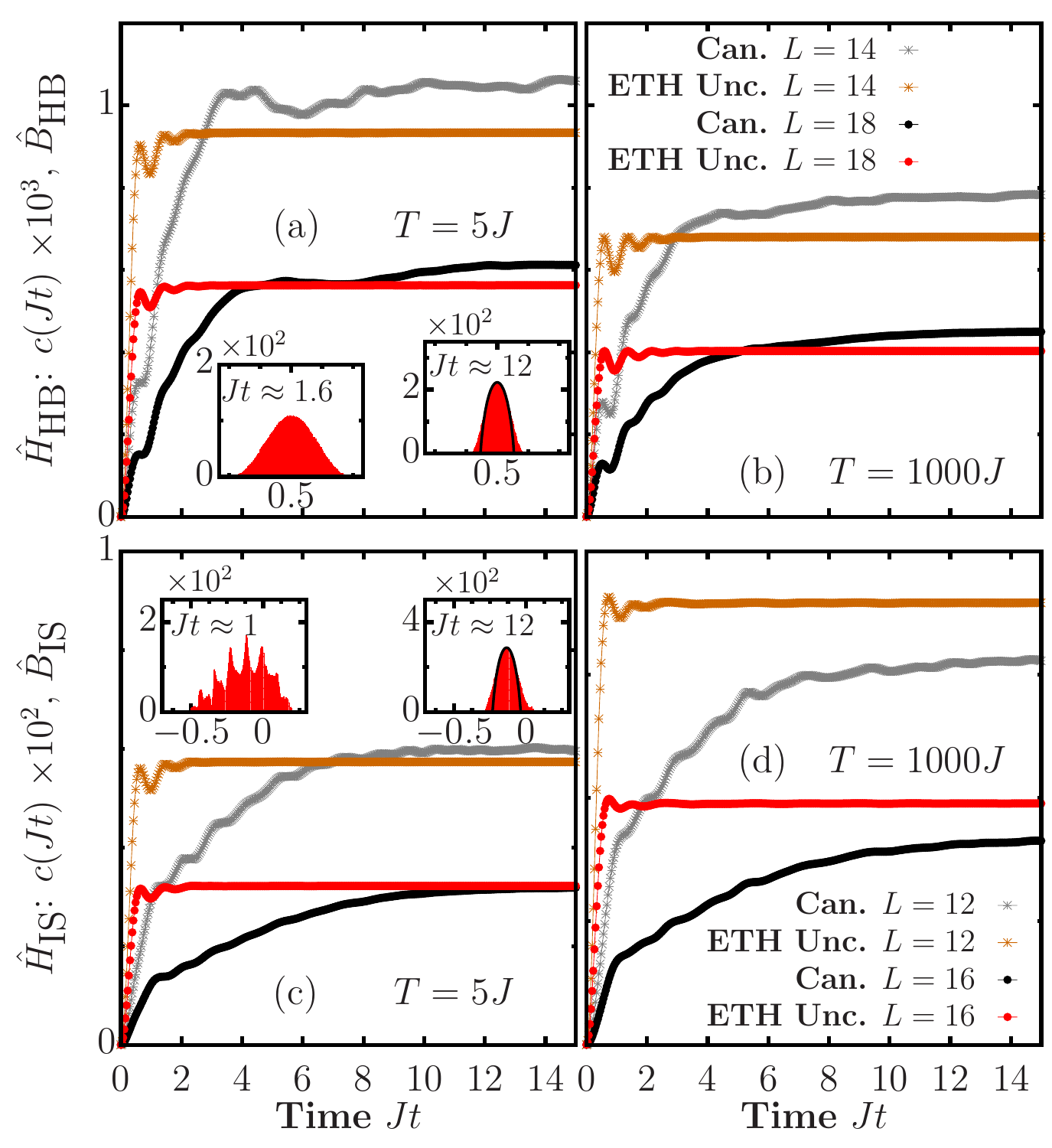}
\caption{Time-dependent square-commutator \eqref{eq_sc} for the operators $\hat{B}_{\textrm{HB}}$ [(a),(b)] and $\hat{B}_{\textrm{IS}}$ [(c),(d)] for $\hat{H}_{\textrm{HB}}$ and $\hat{H}_{\textrm{IS}}$ at temperatures $T = 5J$ [(a) and (c)] and $T = 1000J$ [(b) and (d)]. The expectation value obtained for a canonical state is compared with the one obtained assuming the ETH and uncorrelated $R_{nm}$ for increasing system size $L$. Insets show the distribution of eigenvalues of the matrix $\hat{O}^{\omega_c=2\pi/t}$ \eqref{eq:o_wc} for the largest system size displayed in each case.}
\label{fig:m4}
\end{figure}

The OTOC dynamics in the canonical ensemble are computed by exact diagonalisation, representing $\hat{O}(t)$ as a time-dependent matrix in the Heisenberg picture following the computation of the commutators in Eq.~\eqref{eq_sc}. On the other hand, the dynamical evaluation of $c(t)$ assuming IID Gaussian statistics in the ETH is done by obtaining the two-point functions from $f_{\hat{O}}(\bar{E}, \omega)$. Further details are included in Appendix~\ref{ap:twopoint} and Appendix~\ref{app:otoc}.

The result of this comparison is shown in Fig.~\ref{fig:m4} for sums of local operators. A discrepancy between the two predictions at short times signals that this regime is indeed governed by correlations between the matrix elements. However, the curves saturate to a similar value at longer times, differing in some cases by a small correction that we attribute to energy fluctuations in the canonical ensemble at finite size (see Appendix~\ref{ap:symm_term}), given that these deviations are less prominent for larger system sizes. Fig.~\ref{fig:m4} shows that the time $t_s$ at which saturation occurs qualitatively increases with system size. 

Interestingly, our data suggest that this saturation time is related to the frequency $\omega_{\rm GOE}$ by $t_s\approx 2\pi/\omega_{\rm GOE}$. The procedure employed so far only allows one to study system sizes available to exact diagonalisation techniques. Regardless, we can proceed visually by showing the distribution of eigenvalues of the matrix $\hat{O}^{\omega_c=2\pi/t}$ at different times (insets of Fig.~\ref{fig:m4}).  While at short times the distribution deviates from the GOE prediction, these deviations are strongly reduced when the OTOC nears saturation, approximately leading to semi-circular distributions. This behaviour indicates that the OTOC's long-time dynamics encodes the statistical properties of $R_{nm}$ and the emergence of random-matrix behaviour at low frequencies.

\section{Estimation of the scaling of $\omega_{\rm GOE}$ with system size in the infinite-temperature regime}
\label{sec:scaling}

Our previous results strongly suggests that the frequency scales divided by $\omega_{\rm GOE}$, studied from the spectrum of banded matrices, have a connection to the saturation timescales of the OTOCs, denoted by $t_s$. This connection could be used to estimate the behaviour of $\omega_{\rm GOE}$ as a function of the system size $L$ from the saturation point of the dynamics of the OTOCs. Notice that the saturation time $t_s$ is not related to the relaxation time of two-point correlations, the dephasing time $t_{\varphi}$, which is expected to be an intensive quantity on general grounds~\cite{sachdev2011}. The saturation time $t_s$ is also generically larger than $t_{\varphi}$. This observation is verified by Fig.~\ref{fig:m4}, since $t_{\varphi}$ determines the fast saturation of the OTOC computed according to the uncorrelated approximation in Eq.~\eqref{ETH_Uncorrelated}. 

In our previous calculations, establishing the connection between $t_s$ and $\omega_{\rm GOE}$ entailed the computation of the unitary operator $\hat{U}$ that renders the Hamiltonian diagonal, i.e., $\tilde{H} = \hat{U}^{\dagger} \hat{H} \hat{U}$, where $\tilde{H}$ is a diagonal matrix with the eigenvalues in its entries. This exact diagonalisation procedure is computationally costly due to the rapid increase of $\mathcal{D}$ as a function of the system size. Having established the relation between $\omega_{\rm GOE}$ and $t_s$, we could evaluate the saturation point in the dynamics of the OTOCs using a different approach to provide and estimation of the scaling of $\omega_{\rm GOE}$.

For this purpose, we employ the concept of dynamical quantum typicality~\cite{Goldstein:2006,Popescu:2006,Luitz:2017,Chiaracane:2021,Richter:2018}. In this framework, it is possible to approximate the unitary dynamics of a given system within an equilibrium ensemble from a single pure state $\ket{\psi}$, which is drawn at random from the Haar measure~\cite{Luitz:2017} on an arbitrary basis $\{ \ket{\phi}_k \}_{k = 1}^{\mathcal{D}}$. We start with 
\begin{align}
\ket{\psi} = \hat{R}\sum_{k=1}^{\mathcal{D}}c_{k} \ket{\phi_k}, \quad c_k \defeq a_k + \textrm{i}b_k,
\end{align}
where $\hat{R}$ is an arbitrary operator on the Hilbert space and $a_k$ and $b_k$ are independent random variables drawn from a normal distribution. The averaged expectation value of an operator $\hat{O}$ in the typical state is equivalent to the expectation value computed with respect to a density matrix $\hat{\rho}$, such that $\overline{O} \defeq \overline{\braket{\psi | \hat{O} | \psi}} \approx \textrm{Tr}[\hat{\rho} \hat{O}]$. In this particular case, $\hat{\rho} = \hat{R} \hat{R}^{\dagger}$. It can be shown that the approximation is more accurate as $\mathcal{D}$ increases~\cite{Chiaracane:2021}. 

We now focus on the infinite-temperature regime, in which we can write
\begin{align}
\hat{\rho} = \frac{\mathds{1}}{\mathcal{D}}\quad \textrm{and} \quad \ket{\psi} = \frac{1}{\sqrt{\mathcal{D}}}\sum_{k=1}^{\mathcal{D}}c_k \ket{\phi_k}.
\end{align}
With this procedure, we may approximate the dynamics of $c(t)$ from Eq.~\eqref{eq_sc} in the infinite-temperature regime by
\begin{align}
\label{eq:c_t_s}
c(t) \approx - \left ( \langle \psi | [ \hat O(t), \hat O]^2 | \psi \rangle - \langle \psi | [ \hat O(t), \hat O] | \psi \rangle^2 \right ),
\end{align}
where the approximation becomes more accurate as $L$ is increased. To provide a better approximation for smaller values of $L$ we carry out an averaging procedure using several different random states $\ket{\psi}$. The dynamics is evaluated in the Schr\"odinger picture, using the method of Krylov subspaces to evaluate time-evolved states (see Appendix~\ref{ap:krylov} for further details). 

\begin{figure}[t]
\fontsize{13}{10}\selectfont 
\centering
\includegraphics[width=1\columnwidth]{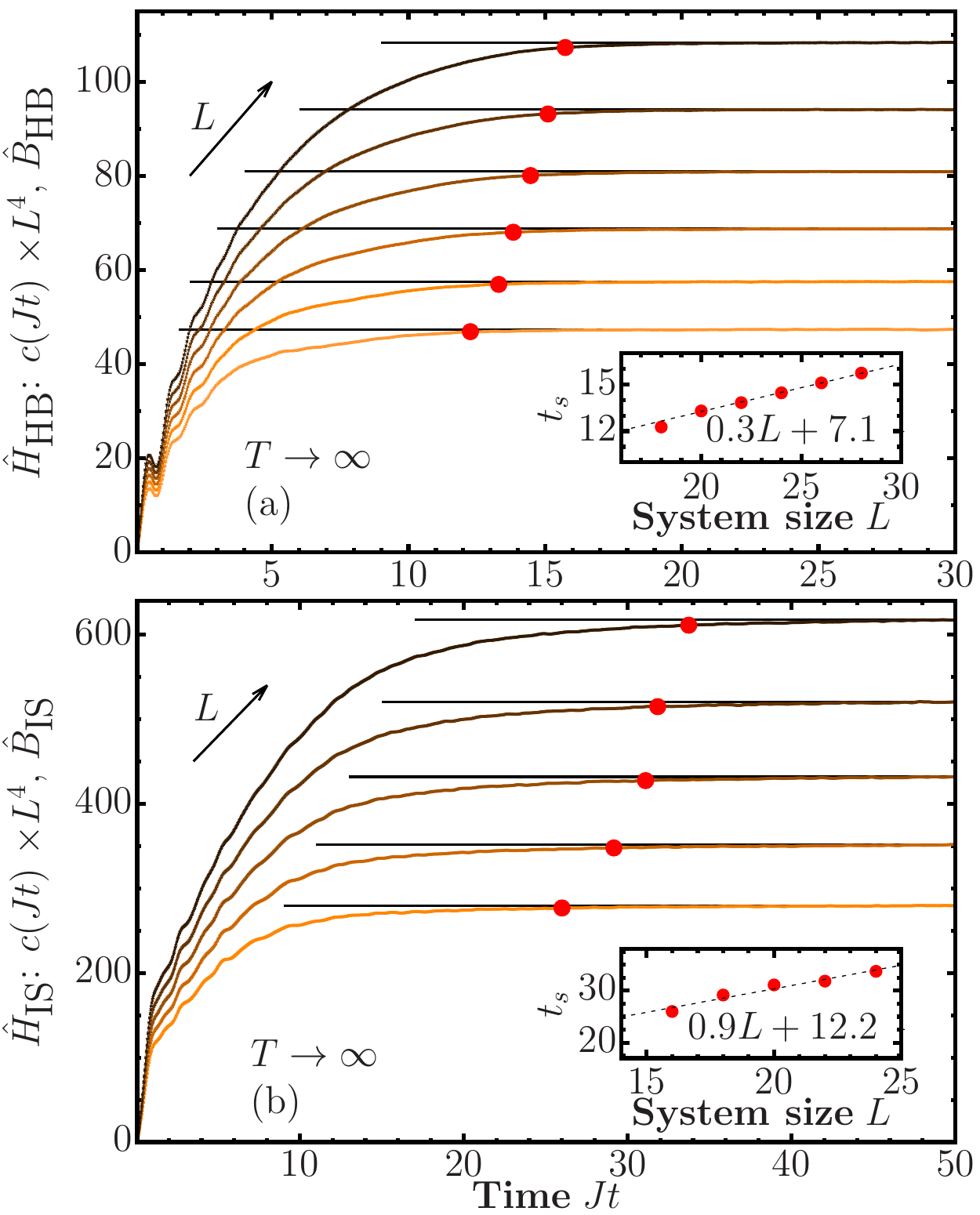}
\caption{Dynamics of the OTOC averaged over many typical states (see Sec.~\ref{sec:scaling}) in the infinite-temperature regime for the (a) $\hat{H}_{\textrm{HB}}$ ($L = 18, \cdots, 28$ increasing for even $L$) and (b) $\hat{H}_{\textrm{IS}}$ ($L = 14, \cdots, 24$ increasing for even $L$) models and their corresponding observables composed of sums of local operators. The circles denote our estimation of the saturation time $t_s$ using the procedure described in Sec.~\ref{sec:scaling}. The insets exhibit the finite-size scaling analysis of the saturation time $t_s$ as a function of $L$, where the dashed lines depict a fit to a linear function $aL+b$.}
\label{fig:m5}
\end{figure}

In Fig.~\ref{fig:m5} we show the results of $c(t)$ averaged over several typical states using the procedure above for the $\hat{H}_{\textrm{HB}}$ [Fig.~\ref{fig:m5}(a)] and the $\hat{H}_{\textrm{IS}}$ [Fig.~\ref{fig:m5}(b)] models for their respective extensive observables. The numerical approach described before allows us to study much larger system sizes (up to $L = 28$ for $\hat{H}_{\textrm{HB}}$ and $L = 24$ for $\hat{H}_{\textrm{IS}}$). The dynamics displayed correspond to the average between many different typical states, which range from $1000$ for the smallest values of $L$, to $2$ for the largest values. The number of realisations is chosen such that the standard deviation along each point in the time trajectory does not surpass $1\%$ of the mean value.

The circles marked in the main panels correspond to the saturation points of the OTOCs. The values of $c(Jt)$ have been scaled by a factor of $L^4$ for visualisation purposes and do not affect the saturation times. To evaluate these saturation values, we first estimate the long-time value of the OTOCs $c(Jt \to \infty)$ from the average of the late dynamics (we use $Jt \in [28,30]$ for $\hat{H}_{\textrm{HB}}$ and $Jt \in [48,50]$ for $\hat{H}_{\textrm{IS}}$). The saturation time is then selected at the value for which $c(Jt_s) = \varepsilon c(Jt \to \infty)$ is reached, where $\varepsilon$ is a certain threshold parameter. The saturation times are highlighted by the circles in the main panels of Fig.~\ref{fig:m5}. There exist some very small finite-size oscillations throughout the dynamics, which introduce a level of uncertainty into the estimation of $t_s$. To account for these, we compare $c(Jt \to \infty)$ against a running-average value in the vicinity of $Jt$. Explicitly so, we compare the long-time estimation of $c(Jt \to \infty)$ against the average within the set $[Jt - 0.5, Jt + 0.5]$ to approximate $t_s$ more accurately. The insets in Fig.~\ref{fig:m5} display the scaling of the saturation time $t_s$ as a function of the system size $L$. In both cases, the saturation time appears to scale linearly with $L$. This behaviour is robust to changes on the parameter $\varepsilon$, as long as $\varepsilon \approx 1$. The displayed $t_s$ results in Fig.~\ref{fig:m5} were obtained with $\varepsilon = 0.99$. 

The observed linear scaling of the saturation time $t_s$, reminiscent of ballistic transport, is qualitatively consistent with the linear front propagation at the butterfly velocity expected in chaotic systems~\cite{shenker2014black,Aleiner16}. This observation and the results obtained for both models would imply that, consequently, $\omega_{\textrm{GOE}} \propto 1/L$ for extensive operators in chaotic many-body systems. The connection of this energy scale to the Thouless energy $\omega_{Th}$, characterising random matrix behaviour in the spectrum of the Hamiltonian, depends on the details of the system. For standard diffusive scaling, $\omega_{\textrm{Th}} \propto 1/L^2 $, one observes $\omega_{\textrm{GOE}} > \omega_{\textrm{Th}}$ (in apparent contrast with Ref.~\cite{Dymarsky2018}, in which strictly local operators were considered). The careful numerical study of the scaling 
of $\omega_{\textrm{Th}}$ with system size required to clarify this issue is beyond the scope of this paper and it is left for future studies.  

\section{Conclusions and outlook}

We have performed a systematic analysis of statistical correlations within the ETH and explored their consequences for the dynamics of quantum information scrambling. Remarkably, we find that correlations between off-diagonal matrix elements indicate the timescale for the onset of random-matrix dynamics in the corresponding OTOC, an experimentally observable quantity. This operator- and temperature-dependent timescale is not apparently connected to hydrodynamic behaviour of linear-response functions, given that the dynamics of stationary two-point correlation functions are independent of statistical matrix-element correlations (see Appendix~\ref{ap:twopoint}). Moreover we have provided an estimation of the scaling of the timescale $t_s$ as a function of the system size $L$, which appears to behave linearly with $L$, consistently with the expected ballistic propagation of combustion-like waves associated to the butterfly effect~\cite{shenker2014black,Aleiner16}. This timescale appears to be connected to the frequency scale $\omega_{\rm GOE} \sim t_s^{-1}$ where random-matrix behaviour is observed from the analysis of banded sub-matrices. The estimation of the scaling as a function of the system size was possible by employing the concept of dynamical typicality in conjunction with computationally optimised Krylov subspace techniques for time evolution. Our results lie at the limit of system size that can be achieved with this numerical approach using parallel algorithms in supercomputers, due to the long timescales required to study saturation of the OTOCs. Finite system size thus remains a limitation to our estimations, despite the fact that the exposed technique allows us to access much larger systems than possible with exact diagonalisation techniques.

\section{Acknowledgements}

We are grateful to L. Vidmar, L. Foini, A. Polkovnikov, J. Richter, A. Dymarsky and Y. Bar Lev for enlightening discussions. M. B. and J. G. acknowledge the DJEI/DES/SFI/HEA Irish Centre for High-End Computing (ICHEC) for the provision of computational facilities, Project No. TCPHY138A, and the Trinity Centre for High-Performance Computing. This work was supported by a SFI-Royal Society University Research Fellowship (J. G.) and the Royal Society (M. B.). J. G. acknowledges funding from European Research Council Starting Grant ODYSSEY (Grant Agreement No. 758403).

\appendix

\setcounter{equation}{0}

\renewcommand{\theequation}{\thesection.\arabic{equation}}

\renewcommand{\thesection}{A\arabic{section}}
\renewcommand{\theHfigure}{Supplement.\thefigure}

\section*{Appendix}

In this appendix, we provide additional information on the numerical computations and on the analytical estimates discussed in the main text. 

In Sec.~\ref{ap:diag}, we study the diagonal matrix elements of local and collective observables in the energy eigenbasis, showing that the two models under analysis obey ETH.
In Sec.~\ref{ap:twopoint}, we consider the real-time dynamics of two point functions and we compare the exact calculation for a thermal state with the result predicted by ETH.  Finite-size corrections are evaluated in Sec.~\ref{ap:symm_term}.
In Sec.~\ref{ap:mean} we study the localisation effects on the eigenvalues distribution of $\hat O^{\omega_c}$ below a given frequency threshold. We follow with Sec.~\ref{app:otoc}, in which we describe the expression of the square-commutator with the assumption of uncorrelated statistics and further details. Finally, Sec.~\ref{ap:krylov} briefly describes the method of Krylov subspaces to evaluate unitary time evolution.

\section{Diagonal matrix elements of observables}
\label{ap:diag}

\begin{figure}[t]
\fontsize{13}{10}\selectfont
\centering
\includegraphics[width=0.93\columnwidth]{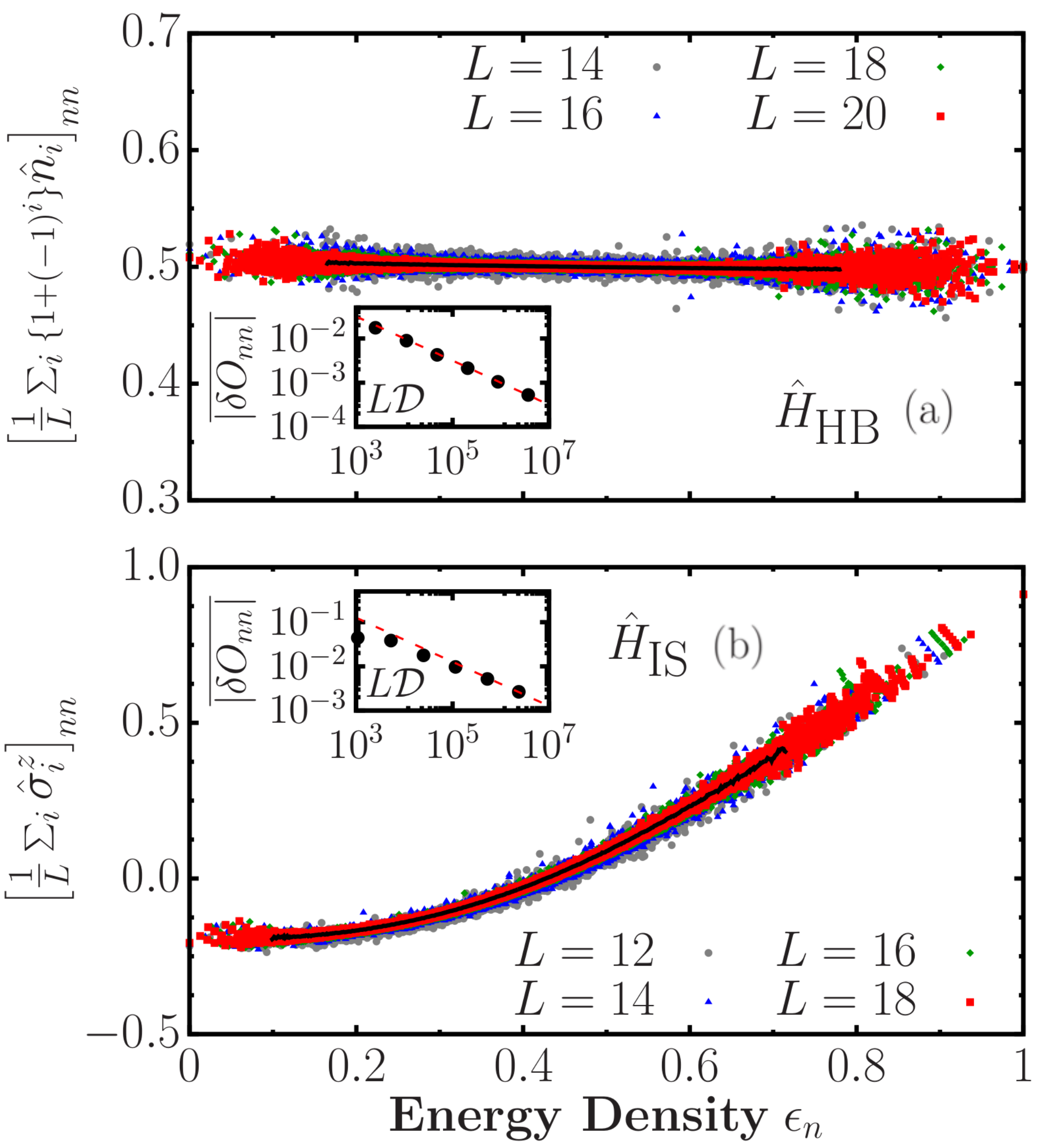}
\caption{Diagonal matrix elements of $\hat{B}_{\textrm{HB}}$ (a) and $\hat{B}_{\textrm{IS}}$ (b) as a function of the energy density $\epsilon_n \defeq (E_n - E_{\textrm{min}}) / (E_{\textrm{max}} - E_{\textrm{min}})$ and of the system size $L$. The black lines depict an approximation of the smooth function $O(\bar{E})$ obtained from a coarse-grained average of the data for the largest system size. The insets show the eigenstate-to-eigenstate fluctuations for different systems sizes, obtained from the eigenvalues in the central region. The dashed lines on the insets show the $(L \mathcal{D})^{-1/2}$ scaling.}
\label{fig:s1}
\end{figure}

A strong indication of eigenstate thermalisation is the behaviour of diagonal matrix elements of observables in the eigenbasis of the Hamiltonian~\cite{Alessio:2016,BrenesLocal2020,Leblond:2019}. In Fig.~\ref{fig:s1} we show the diagonal matrix elements of $\hat{B}_{\textrm{HB}}$ [panel (a)] and of $\hat{B}_{\textrm{IS}}$ [panel (b)] for the non-integrable models studied in this work. We defined the energy density $\epsilon_n \defeq (E_n - E_{\textrm{min}}) / (E_{\textrm{max}} - E_{\textrm{min}})$ and computed all the matrix elements in the eigenbasis of the Hamiltonian by full diagonalisation. It can be observed that, as the system size $L$ is increased, the support over which the matrix elements exist shrink. This observation strongly suggests that in the thermodynamic limit, the diagonal matrix elements can be described by a smooth function $O(\bar{E})$ corresponding to the microcanonical prediction (note that the second term in Eq.~\eqref{eq:eth} is exponentially suppressed in Hilbert space dimension $\mathcal{D}$). The black lines in Fig.~\ref{fig:s1} depict an approximation of the smooth function $O(\bar{E})$, obtained from a coarse-grained average of the data for the largest system size. 

The insets in Fig.~\ref{fig:s1} highlight the trend of the absolute value of the eigenstate-to-eigenstate fluctuations, defined as
\begin{align}
\overline{|\delta O_{nn} |} \defeq \overline{| O_{nn} - O_{n+1 n+1}|},
\end{align}
computed for 20\% of the total eigenvalues in the centre of the spectrum. The dashed line on the insets in Fig.~\ref{fig:s1} corresponds to the scaling $L\mathcal{D}^{-1/2}$, expected in the high-temperature regime of non-integrable models. We remark that the eigenstate-to-eigenstate fluctuations for sums of local observables scales like $L\mathcal{D}^{-1/2}$, as opposed to the more common $\mathcal{D}^{-1/2}$ scaling observed for local observables. This behaviour can be attributed to the $1 / \sqrt{L}$ scaling of the Schmidt norm for this class of observables~\cite{Vidmar:2019,Vidmar2:2020,Leblond:2019, BrenesLocal2020}.

The results shown in Fig.~\ref{fig:s1} indicate that the ETH is obeyed by the models and observables considered in the main text for the parameters selected, away from non-generic features observed at the edges of the spectrum. 

\section{Dynamics of two-point correlation functions}
\label{ap:twopoint}

\begin{figure*}[t]
\fontsize{13}{10}\selectfont 
\centering
\includegraphics[width=1.98\columnwidth]{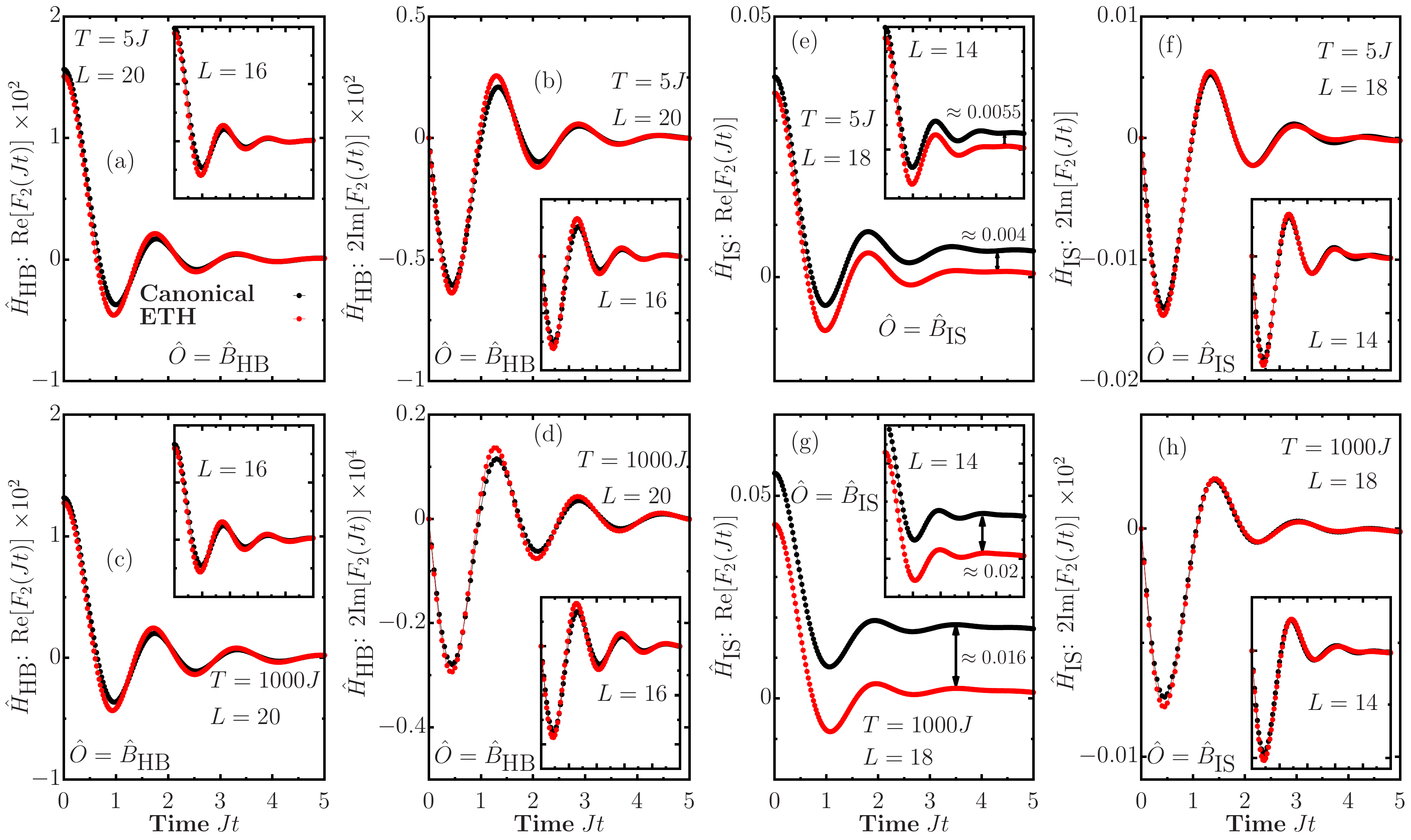}
\caption{Dynamics of the two-point correlation function evaluated in the canonical ensemble at temperature $T$ and in the ETH with a compatible energy density for sums of local operators. In [(a)-(d)] we show the results for $\hat{H}_{\textrm{HB}}$ and in [(e)-(h)] the corresponding results for $\hat{H}_{\textrm{IS}}$ at different temperatures for different system sizes as highlighted in the figure.}
\label{fig:s2}
\end{figure*}

The ETH in Eq.~\eqref{eq:eth} provides the form of the off-diagonal matrix elements required to predict the dynamics of two-point correlation functions at thermal equilibrium. We are interested in correlation functions of the form
\begin{align}
\label{eq:f2}
F_2(t) \defeq \langle \hat{O}(t) \hat{O}(0) \rangle_c \defeq \langle \hat{O}(t) \hat{O}(0) \rangle - \langle \hat{O}(t) \rangle \langle \hat{O}(0) \rangle,
\end{align}
where the expectation values are evaluated in one of the ensembles of statistical mechanics. One could, for instance, consider the canonical ensemble. For an operator $\hat{A}$ in such case, we have that $\langle \hat{A} \rangle = \textrm{Tr}[\hat{\rho} \hat{A}]$, where $\hat{\rho} = e^{-\beta \hat{H}} / Z$ is the density matrix operator for a system with Hamiltonian $\hat{H}$, partition function $Z = \textrm{Tr}[e^{-\beta \hat{H}}]$ and inverse temperature $\beta = 1 / T$. In Eq.~\eqref{eq:f2}, the operators are written in the Heisenberg picture $\hat{O}(t) = e^{\textrm{i} \hat{H} t} \hat{O}(0) e^{-\textrm{i} \hat{H} t}$.

On the other hand, eigenstate thermalisation suggests that such expectation values could be evaluated for a {\em single eigenstate} $\ket{E_n}$. The evaluation is simpler if one instead considers the symmetric and anti-symmetric response functions which yield, respectively, the real and imaginary parts of $F_2(t)$. Written in such fashion we have
\begin{align}
S^{+}_{\hat{O}}(E_n, t) &\defeq \braket{E_n | \{ \hat{O}(t), \hat{O}(0) \} | E_n}_c = 2\,\textrm{Re}[F_2(E_n, t)]\nonumber \\
S^{-}_{\hat{O}}(E_n, t) &\defeq \braket{E_n | \,[ \hat{O}(t), \hat{O}(0) \,] | E_n}_c = 2\textrm{i}\, \textrm{Im}[F_2(E_n, t)],
\end{align}
where $\{\cdot, \cdot \}$ and $[ \cdot, \cdot ]$ stand for the anti-commutator and commutator, respectively. In this notation, $F_2(t)$ is the one considered in the canonical ensemble, while $F_2(E_n, t)$ is the one evaluated for a single energy eigenstate. 

Following the standard derivation \cite{Srednicki:1999,Alessio:2016,Vidmar:2020} from the ETH in Eq.~\eqref{eq:eth}, one obtains the correlation functions in frequency domain in the thermodynamic limit
\begin{align}
\label{eq:f2_w_re_im}
S^{+}_{\hat{O}}(E_n, \omega) &\approx 4\pi \cosh(\beta \omega /2) |f_{\hat{O}}(E_n, \omega)|^2, \nonumber \\
S^{-}_{\hat{O}}(E_n, \omega) &\approx 4\pi \sinh(\beta \omega /2) |f_{\hat{O}}(E_n, \omega)|^2.
\end{align}
Given that these relations are symmetric and anti-symmetric, respectively, their Fourier transforms to yield the correlation functions in the time domain are simplified to
\begin{align}
\label{eq:f2_t_re_im}
\textrm{Re}[F_2(E_n, t)] = \int_0^{\infty} d\omega \cos(\omega t) S^{+}_{\hat{O}}(E_n, \omega), \nonumber \\
\textrm{Im}[F_2(E_n, t)] = \int_0^{\infty} d\omega \sin(\omega t) S^{-}_{\hat{O}}(E_n, \omega).
\end{align}
At this point is important to make two observations with respect to Eq.~\eqref{eq:f2_t_re_im}. First, in the thermodynamic limit we expect $F_2(E_n, t) = F_2(t)$. This immediately follows from the association of the energy $E_n$ to a corresponding canonical inverse temperature $\beta$ by assigning $E_n = \langle E \rangle = \textrm{Tr}[\hat{H} e^{-\beta \hat{H}}] / Z$. Second, in Eq.~\eqref{eq:f2_w_re_im}, there is no dependency of the random variable $R_{nm}$. This follows from the fact that this term enters the dynamical correlations in the form of the average of $|R_{nm}|^2$, which is unity by assumption~\cite{Srednicki:1999,Alessio:2016}. Indeed, it suffices that this random variable has a well-defined variance for the $|R_{nm}|^2$ term to vanish from the final expressions.  

The equivalency between the dynamics of two-point correlation functions in statistical mechanics and the corresponding dynamics of the same object predicted by the ETH can be observed in generic systems. Following Eqs.~\eqref{eq:f2_w_re_im} and \eqref{eq:f2_t_re_im}, the dynamics of the correlation functions depend solely on the function of $f_{\hat{O}}(E_n, \omega)$; which is, in general, system- and observable-dependent. A commonly-used procedure~\cite{Mondaini2017,Khatami2013,BrenesFisher2020,BrenesLocal2020,BrenesFreq2020,Vidmar:2019} to isolate this function in generic systems involves a frequency-resolved analysis of the matrix elements of a given observable in the energy eigenbasis. One focuses on a small window of energies and extracts the off-diagonal matrix elements of an operator $\hat{O}$ in the eigenbasis of the Hamiltonian. For finite-size systems, the fluctuations present are accounted for by considering not a single eigenstate, but a collection of eigenstates around a given energy $E_n$. A coarse-grained average then leads to a smooth function $e^{-S(E_n)/2}f_{\hat{O}}(E_n, \omega)$. The entropy term, $e^{-S(E_n)/2}$, is not a function of $\omega$ and in principle needs to be evaluated. Instead of evaluating this term directly, we first compute the symmetric correlation function $S^{+}_{\hat{O}}(E_n, \omega)$ and normalise it by the sum rule
\begin{align}
\int_{-\infty}^{\infty} d\omega S^{+}_{\hat{O}}(E_n, \omega) = 4\pi \left[ \braket{E_n | \hat{O}^2 | E_n} - \braket{E_n | \hat{O} | E_n}^2 \right],
\end{align}
while the anti-symmetric correlation function $S^{-}_{\hat{O}}(E_n, \omega)$ follows from Eq.~\eqref{eq:f2_w_re_im}, which is the manifestation of the fluctuation-dissipation theorem. 

This procedure can be applied to the models and observables described in the main text. In Fig.~\ref{fig:s2}, we show the dynamics of both the real and imaginary parts of the two-time correlation function in Eq.~\eqref{eq:f2}, for sums of local observables [Eq.~\eqref{eq:obs_ext}]. The dynamics of the two-point correlation function in the canonical ensemble were evaluated by direct diagonalisation of the propagator $e^{-\textrm{i}\hat{H}t}$ acting on the density operator $\hat{\rho}$, while the dynamics from the ETH were evaluated using the procedure described above. For the latter, we computed $e^{-S(E_n)/2}f_{\hat{O}}(E_n, \omega)$ by considering a target energy $\bar{E} = \textrm{Tr}[\hat{\rho} \hat{H}]$ consistent with the canonical temperature $T$ and averaging all the off-diagonal matrix elements within an energy window of width $0.075\epsilon$, where $\epsilon \defeq E_{\textrm{max}} - E_{\textrm{min}}$ [see Refs.~\cite{BrenesFisher2020,BrenesLocal2020,Khatami2013,Mondaini2017} for further details on the extraction of $f_{\hat{O}}(E_n, \omega)$].  

For finite-size systems, the connected symmetric correlation function contains a time-independent term that is not present if one evaluates the same object on a single eigenstate. This term is expected to vanish in the thermodynamic limit, see Sec.~\ref{ap:symm_term} for an extended discussion. The difference stems from the fluctuations in the canonical ensemble, a term which is already small for $\hat{B}_{\textrm{HB}}$ in Fig.~\ref{fig:s2}[(a),(c)], but not as much for $\hat{B}_{\textrm{IS}}$ in Fig.~\ref{fig:s2}[(e),(g)]. The difference, however, as highlighted in Fig.~\ref{fig:s2}[(e),(g)], becomes smaller as the system size is increased. The dynamics observed strongly suggest that such seemingly-constant discrepancy can be attributed to finite-size corrections.

Though the fine details and the actual form of the decay of $F_2(t)$ depend on the system being considered, it can be observed from Fig.~\ref{fig:s2} that $F_2(E_n, t) \approx F_2(t)$, an approximation that becomes more accurate as the thermodynamic limit is approached. It is important to remark that such prediction is accurate not only at high temperature (bottom panels in Fig.~\ref{fig:s2}), but at finite temperature (top panels in Fig.~\ref{fig:s2}) as well.   

\section{Finite-size term in symmetric correlation functions}
\label{ap:symm_term}

The symmetric correlation function evaluated on a thermal state $F_2(t)$ [Eq.~\eqref{eq:f2}] differs from the one computed on a single eigenstate $F_2(E_n, t)$ by a sub-leading term in the thermodynamic limit \cite{Alessio:2016}.

The expectation value computed in the canonical ensemble is defined as $\langle \cdot \rangle \defeq \Tr[\hat \rho \, \cdot ] = \sum_n p_n \bra{E_n} \cdot \ket{E_n}$, where in the case of a canonical density matrix one has  $p_n = e^{-\beta E_n}/Z$. In general, other ensembles can be considered, provided that the distribution of the $p_n$ is sufficiently peaked around some average energy $E=\braket{\hat H}$ with small variance $\delta E^2= \braket{\hat H^2}-\braket{\hat{H}}^2$, i.e. $\delta E^2 / E^2\sim 1/L$. \\
Defining $O_{nm} \defeq \bra{E_n} \hat O \ket{E_m}$, the two-point function reads
\begin{align}
\label{eq:F2Exp}
        F_2(t) 
        & = \sum_{nm} p_n e^{-\textrm{i}(E_m-E_n)t}\, O_{nm} O_{mn} - \left (\sum_n p_n O_{nn}\right )^2 
        \nonumber \\
        & = \sum_{n\neq m} p_n e^{-\textrm{i}(E_m-E_n)t}\, O_{nm} O_{mn} + \braket{\hat{O}^2} - \braket{\hat{O}}^2 
        \nonumber \\
        & = \sum_n p_n \, \bra{E_n}\hat O(t) \hat O \ket{E_n}_c\, + \delta \hat{O}^2 \ ,
\end{align}
where in the second line we have identified $\langle \hat{O}^2\rangle=\sum_n p_n [O_{nn}]^2$,  $\langle \hat{O} \rangle=\sum_n p_n O_{nn}$ and defined $\delta \hat{O}^2 \defeq \braket{\hat{O}^2} - \braket{\hat{O}}^2$. 
The first term in Eq.~\eqref{eq:F2Exp} coincides with the two-point function evaluated on a single eigenstate $F_2(E_n, t)$, while the second one is a time-independent quantity that can be shown to be sub-leading, i.e.,
\[
F_2(t) \sim F_2(E, t) + \mathcal O(1/L) \quad \text{for}  \quad L\gg 1 \ .
\]

Using the ETH in Eq.~\eqref{eq:eth} and the fact that $p_n$ is peaked around energy $E$, one can write down a Taylor expansion around $E$ and the diagonal elements $O_{nn} = O(E_n)$ as
\begin{equation}
    O(E_n) = O(E) + (E_n-E) O'(E) + \frac {(E_n-E)^2} 2 O''(E) + \dots \ ,
\end{equation}
where $O'(E)$ and $O''(E)$ are respectively the first and second derivatives with respect to energy of the microcanonical smooth function $O(E)$ in Eq.~\eqref{eq:eth}. It then follows that, to leading order,
\begin{equation}
    \delta O^2 = \left( \frac{\partial O}{\partial E}\right)^2 \delta E^2.
\end{equation}
We require $\delta E^2 / E^2\sim 1/L$, then $\delta O^2$ is sub-leading in the thermodynamic limit. For finite-size systems, however, this term corresponds to the time-independent finite-size correction discussed in the previous section, observed in Fig.~\ref{fig:s2}.

\section{Inferring GOE- and Poisson-distributed frequency regimes from the mean ratio of level spacings}
\label{ap:mean}

\begin{figure}[b]
\fontsize{13}{10}\selectfont 
\centering
\includegraphics[width=1\columnwidth]{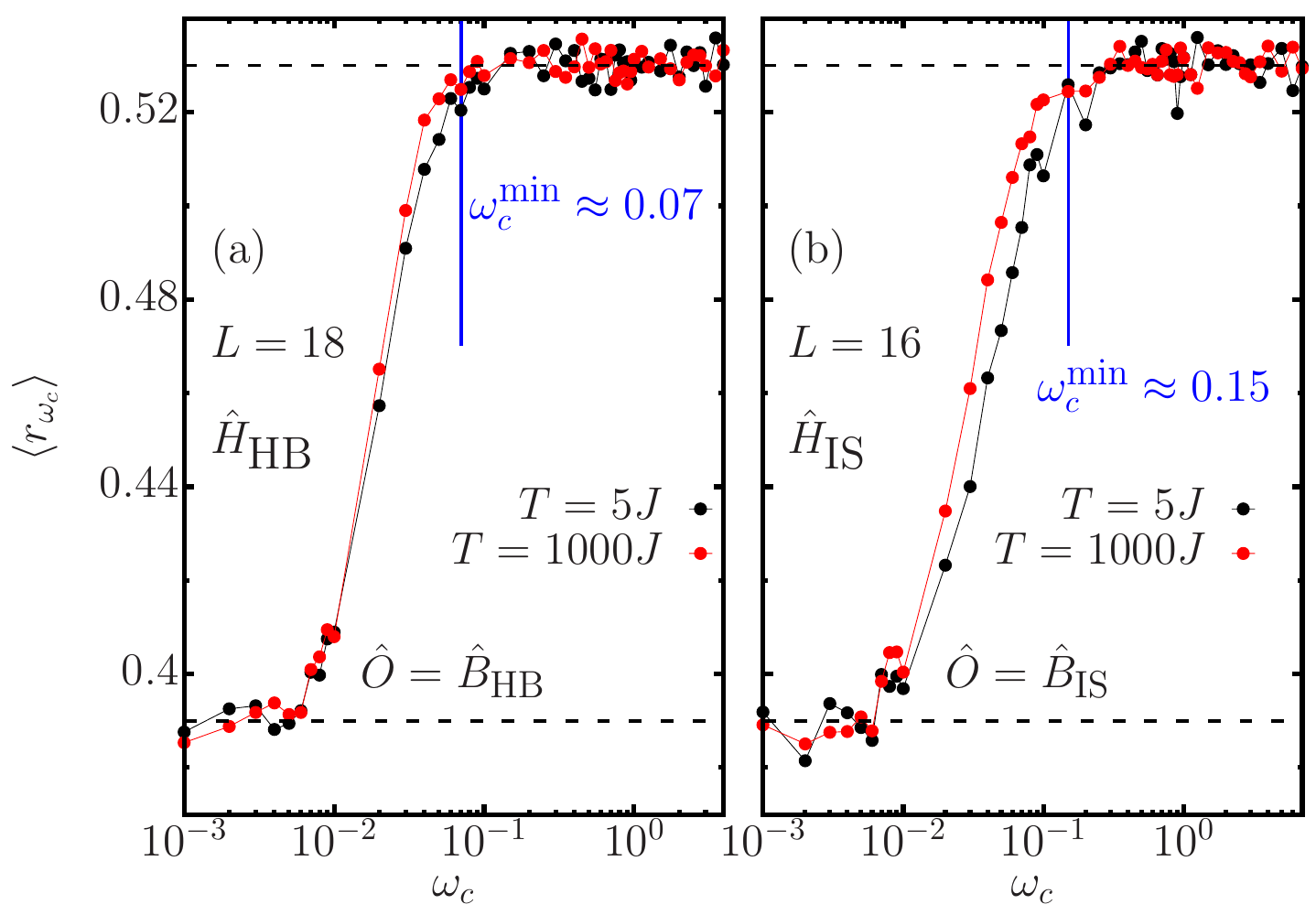}
\caption{Mean ratio of adjacent level spacings $\langle r_{\omega_c} \rangle$ as a function of the cutoff frequency $\omega_c$ for the (a) $\hat{H}_{\textrm{HB}}$ and (b) $\hat{H}_{\textrm{IS}}$ models, at fixed system size for two different values of temperature $T = 5J$ and $T = 1000J$.}
\label{fig:s3}
\end{figure}

Throughout this work, we considered banded sub-matrices to determine the degree of correlations within the statistical matrix $R_{nm}$. From Eq.~\eqref{eq:o_wc}, $\omega_c$ determines the frequency value associated to a given banded sub-matrix. We considered $\omega_c$ only above a given threshold, due to the fact that below this threshold the eigenvalues of $\hat{O}^{\omega_c}$ become uncorrelated because of localisation effects~\cite{Richter2020}. In this frequency regime, the adjacent eigenvalue level spacings of $\hat{O}^{\omega_c}$ are Poisson-distributed. 

The relevant frequency regime in our work is the one dictated by the largest $\omega_c$ where the eigenvalues $\lambda^{\omega_c}_{\alpha}$ are still uncorrelated. This implies that there are resonant timescales $t \sim 2\pi / \omega_c$ for which the dynamics are dictated by uncorrelated energy modes. In Ref.~\cite{Richter2020}, it was shown that the eigenvalues of $\hat{O}^{\omega_c}$ are uncorrelated even in the regime where the distribution of level spacings follows the GOE. For these reasons, it is important to restrict ourselves to values of $\omega_c$ for which the eigenvalues of $\hat{O}^{\omega_c}$ are uncorrelated and the level spacings follow the GOE. These regimes can be probed by studying the mean ratio of adjacent level spacings, defined as
\begin{align}
\braket{r_{\omega_c}} \defeq \frac{1}{M}\sum_{\alpha} \frac{\textrm{min}\{ \Delta_{\alpha}, \Delta_{\alpha + 1} \}}{\textrm{max}\{ \Delta_{\alpha}, \Delta_{\alpha + 1} \}},
\end{align}
where $\Delta_{\alpha} = | \lambda_{\alpha + 1}^{\omega_c} - \lambda_{\alpha}^{\omega_c}|$. We performed the average over all adjacent level spacings, i.e., $M \approx D^{\prime}$. We have that $\braket{r_{\omega_c}} \approx 0.53$ for a distribution following the GOE and $\braket{r_{\omega_c}} \approx 0.39$ for Poisson-distributed random variables.  

In Fig.~\ref{fig:s3}(a) we show the mean ratio of adjacent level spacings for the $\hat{H}_{\textrm{HB}}$ model as a function of $\omega_c$. It can be observed that the value of $\omega_c$ for which the onset of the GOE is observed is rather similar between the values of temperature $T = 5J$ and $T = 1000J$ chosen. We do not expect this behaviour to be generic. On the contrary, the onset of the GOE is typically observed at different values of $\omega_c$ for different system sizes $L$ and observables. To avoid the aforementioned localisation effects, we restricted our analyses to the values of $\omega_c$ above the $\omega_c^{\textrm{min}}$ denoted in Fig.~\ref{fig:s3}. For all $\omega_c > \omega_c^{\textrm{min}}$, the associated sub-matrices exhibit a value of $\langle r_{\omega_c} \rangle \approx 0.53$. This, however, does not imply that the matrix elements display correlations or lack thereof. As we have shown in the main text, there exist a regime for which correlations build up as $\omega_c$ is increased above $\omega_c^{\textrm{min}}$. Similar results are observed for the $\hat{H}_{\textrm{IS}}$ model in Fig.~\ref{fig:s3}(b), with the only difference noticed at the specific $\omega_c^{\textrm{min}}$ values for which the onset of the GOE mean ratio of level spacings is obtained. 

\section{Out-of-time-order correlators}
\label{app:otoc}

In this section, we evaluate the dynamics of four-point functions to investigate whether correlations translate into an observable effect. We will start by computing the expectation values of time-dependent correlation functions on a single eigenstate, while neglecting any effect of correlations. We will then proceed to evaluate the same objects in the canonical ensemble. The difference in the dynamics will allow us to probe the effect of the correlations exposed in the main text.

\subsection{Four-point functions within Gaussian statistics}
\label{app:otocDeriva}

In Ref.~\cite{Murthy19}, the ETH result for the thermally-regulated OTOC was computed. It was done by interlacing fractional powers of the density matrix with each operator. The case studied in Ref.~\cite{Murthy19} had a feature for which the diagonal expectation values vanish, i.e., $O_{nn}=O(E_n)=0$.

In this work, we considered multi-time correlation functions evaluated on standard thermal states and we generalised the result for cases in which $O_{nn}$ is non-zero.
Consider the following four point connected correlator
\begin{align}
\begin{split}
\label{eq:defF4c}
F_c(t_1, t_2, t_3, t_4) & \defeq
\braket{ \hat O(t_1)\, \hat O(t_2)\,\hat O(t_3)\,\hat O(t_4) } \\
& \quad - \braket{ \hat O(t_1) \,\hat O(t_2)}\,\braket{\hat O(t_3)\,\hat O(t_4) }
\end{split}
\end{align}
where all operators are written in the Heisenberg representation, $\hat O(t)=e^{i\hat Ht}\hat Oe^{-i\hat Ht}$.
All time-ordered and out-of-time-ordered correlation functions can be constructed from $F_c$ with a suitable choice of arguments. In particular, we focus on the standard OTOC
\begin{equation}
\label{eq:Fotoc}
F_{\text{OTO}}(t)  \defeq 
 \braket{ \hat O(t) \hat O  \hat O(t) \hat O }  - \braket{ \hat O(t) \hat O }^2 =  F_c( t, 0, t, 0),
\end{equation}
and the square commutator
\begin{align}
\label{eq_scApp}
c(t) & \defeq - \left ( \langle [ \hat O(t), \hat O]^2 \rangle - \langle [ \hat O(t), \hat O]\rangle^2 
\right ) \\
& \quad= F_c( t, 0, 0, t)+F_c( 0, t, t, 0)- 2 \text{Re} F_c( t, 0, t, 0)\ . \nonumber
\end{align}

We now restrict our analysis with the assumption that the matrix elements in the ETH from Eq.~\eqref{eq:eth} are uncorrelated Gaussian variables, i.e. 
\begin{subequations}
\label{eq:Rexp}
\begin{align}
\label{eq:R2exp}
\overline{R_{\a\be}\;R_{\g\de}} & = \delta_{\a\de}\;\delta_{\be\g} \\
\label{eq:R4exp}
\begin{split}
\overline{R_{\a\be}\;R_{\be\g}\;R_{\g\de}\;R_{\de\a}} 
& = 
\overline{R_{\a\be}\;R_{\be\g}} \; \overline{R_{\g\de}\;R_{\de\a}} 
\\ \quad 
&+ \overline{R_{\a\be}\;R_{\g\de}} \; \overline{R_{\be\g}\;R_{\de\a}} \\
&+ \overline{R_{\a\be}\;R_{\de\a}} \; \overline{R_{\be\g}\;R_{\g\delta}} \ .
\end{split}
\end{align}
\end{subequations}
This allows us to re-write the four point function Eq.~\eqref{eq:defF4c} evaluated over $\hat \rho=\sum_n p_n \ket{E_n}\bra{E_n}$ as
\begin{equation}
F_c(t_1, t_2, t_3, t_4) = \sum_n p_n \, F_c(E_n, t_1, t_2, t_3, t_4) \ ,
\end{equation}
where $F_c(E_n, t_1, t_2, t_3, t_4)$ is the micro-canonical expectation value, i.e., Eq.~\eqref{eq:defF4c} computed over a single eigenstate $\ket{E_n}$. One can observe that the same result holds from a purely out-of-equilibrium calculation, where the expectation value in Eq.~\eqref{eq:defF4c} is taken over an initial pure state $\ket{\Psi}$. In this case, the distribution of the $p_n$ is given by the overlaps with the initial state $p_n = |\langle E_n|\Psi\rangle|^2$.
Similar to the procedure undertaken in Sec.~\ref{ap:symm_term}, one can show that if $p_n$ is compatible with statistical mechanics --- with average energy $E$ and small variance $\delta E^2/E^2\sim 1/L$ --- the leading term of Eq.~\eqref{eq:defF4c} is given by the single-eigenstate expectation value. One can, however, expect finite size effects, as shown in Fig.~\ref{fig:s2}. In the following, we consider only the microcanonical single eigenstate four-point functions and omit $E_n$ in the notations.

Using Eq.~\eqref{eq:Rexp}, one can re-write the four-point function in Eq.~\eqref{eq:defF4c} directly in terms of the two-point functions from Eq.~\eqref{eq:f2} as
\begin{widetext}
\begin{align}
\label{eq:result}
\begin{split}
F_c(t_1, t_2, t_3, t_4) & = f_1(t_1, t_2, t_3, t_4) + c_1(t_2-t_3) + c_2(t_1-t_4, t_2-t_3) \ ,
\end{split}
\end{align}
with
\begin{align}
\label{eq:c1}
c_1(t_2-t_3) &= O'^2 F_2''(t_2-t_3),
\end{align}
\begin{align}
\label{eq:c2}
c_2(t_1-t_4, t_3-t_2) & = 
F_2(t_1-t_4) \;F_2(t_2-t_3)  + F'_2(t_1-t_4) \;\frac{\partial F_2(t_2-t_3) }{\partial E},
\end{align}
and
\begin{align}
\label{eq:f1}
f_1(t_1, t_2, t_3, t_4)  &=  O^2\left [F_2(t_1-t_3) + F_2(t_2-t_4)\right . 
+\left . F_2(t_1-t_4) + F_2(t_2-t_3) \right ] \\
& +
O O' \left [F'_2(t_1-t_3) + F'_2(t_2-t_4) + 2F'_2(t_2-t_3) \right ]  + \frac 12 O O'' \left [F''_2(t_1-t_3) + F''_2(t_2-t_4) + 2F''_2(t_2-t_3) \right ] \nonumber,
\end{align}
\end{widetext}

\noindent where $F_2(t)$ is written as defined in Eq.~\eqref{eq:f2} $O\,$, $O\,'$ and $O\,''$ are obtained from the diagonal expectation value of the operator $\hat{O}$ and its first and second derivative with respect to energy evaluated at $E_n$.

The functions $F'_2(t)$  and $F''_2(t)$ can be written as
\begin{subequations}
\label{eq:F22Def}
\begin{align}
\label{f2primo}
2 \text {Re} F'_2(t) & \equiv -  \sum_{\beta\neq n} \omega_{n \beta} (e^{\textrm{i} \omega_{n \beta} t} +e^{-\textrm{i} \omega_{n \beta} t} )\, |f_{n \beta}|^2\, e^{-S_{n \beta}} \nonumber\\
& = -  \frac 1{2\pi}\, \int d\omega \; \omega\; S_{\hat O}(E_n, \omega) e^{\textrm{i}\omega t} \ ,\\
%%%%%
\label{f2secondo}
2 \text {Re}F''_2(t) & \equiv   \sum_{\beta\neq n} \omega_{n \beta}^2 (e^{\textrm{i} \omega_{n \beta} t} +e^{-\textrm{i} \omega_{n \beta} t} )\, |f_{n \beta}|^2\, e^{-S_{n \beta}} \nonumber \\
& =  \frac 1{2\pi}\, \int d\omega \;\omega^2\; S_{\hat O}(E_n, \omega) e^{\textrm{i}\omega t}\ ,
\end{align}
\end{subequations}
where in the second line of Eq.~\eqref{eq:F22Def} we have identified sums with integrals and expanded the entropy terms around energy $E_n$. The resulting $ S_{\hat O}(E_n, \omega)$ is the symmetric response function which can be computed within ETH as stated in Eq.~\eqref{eq:f2_w_re_im}. Note that $f_1$ and $c_1$ vanish when $O(E_n)=0$, and do not contain information about the distribution of the $R_{n \beta}$ in the ETH. 

To derive Eq.~\eqref{eq:result}, we first write down $F_c$ in the energy eigenbasis and, after invoking the ETH in Eq.~\eqref{eq:eth} and the Gaussian uncorrelated approximation in Eq.~\eqref{eq:Rexp}, we express $O_\be=O(E_\be)$ as a Taylor expansion around the energy $E_n$ at second order in frequencies $\omega_{n\beta}$. Then, upon substituting discrete sums with integrals, we expand the exponential of entropy terms as done before in Eqs.~\eqref{eq:F22Def}, to obtain the final result in Eq.~\eqref{eq:result}.

Therefore, the OTOC and square-commutator can be directly written as combinations of $f_1, c_1$ and $c_2$ [Eqs.~\eqref{eq:c1}, \eqref{eq:c2} and \eqref{eq:f1}] as
%\begin{align}
%F_{\text{OTO}}(t) & =  \; f_1(t) +  \; c_1(t) + \; c_2(t) \\
%\label{eq:ct_eth}
%c(t)  & = 2\left[ c_1(0) + c_2(0) - c_1(t) - c_2(t) \right] \ .
%\end{align}
\begin{align}
F_{\text{OTO}}(t) & =  \; f_1(t, 0, t, 0) +  \; c_1(t) + \; c_2(t, -t) \\
\label{eq:ct_eth}
c(t)  & = 2\left[ c_1(0) + c_2(0, 0) - c_1(t) - c_2(t, -t) \right] \ .
\end{align}
Within this approximation, the leading terms in the system size of the square-commutator and of the OTOC read
\begin{align}
\label{eq:cteth}
    F_{\text{OTO}} & = |F_2(t)|^2 + 2\, O(E_n)^2\,  
    \Re{\left [ F_2(t)+F_2(0)\right ]} \ , \\
    c(t) & = 2 |F_2(0)|^2 - 2 |F_2(t)|^2 \ .
\end{align}
In fact, all the terms containing derivatives with respect to energy are proportional to $1/L$ and are therefore sub-leading in $L$ with respect to $|F_2(t)|^2$, both for local or sums of local operators.
Note, however, that these terms could be relevant in large-$L$ chaotic models at intermediate times, where the square-commutator is expected to grow exponentially as $\sim e^{2\lambda t}/L^2$, with $\lambda$ the classical Lyapunov exponent \cite{Pappalardi2020Quantum}.

The square commutator as defined in Eq.~\eqref{eq_scApp} can be directly evaluated from the ETH assuming uncorrelated $R_{nm}$ from Eq.~\eqref{eq:eth} using the procedure described above. The resulting expression depends only on the dynamics of two-point functions which can be evaluated using the procedure described in Sec.~\ref{ap:twopoint}.

\section{Method of Krylov subspaces for time evolution}
\label{ap:krylov}

The dynamics of $c(t)$ in Eq.~\eqref{eq:c_t_s} can be approximated accurately in the Schr\"odinger picture by the method of Krylov subspaces. The idea is to evaluate the action of the propagator onto a pure state to obtain a time-evolved state, i.e., $\ket{\psi(t)} = e^{-\textrm{i}\hat{H}t}\ket{\psi(0)}$. We address this by employing a computationally-optimised approach to the method of Krylov subspaces. With this method, we evaluate $\ket{\psi(t)}$ by computing the action of $e^{-\textrm{i}\hat{H}t}$ onto $\ket{\psi(0)}$. This is done by a polynomial approximation to $\ket{\psi(t)}$ from within the Krylov subspace
\begin{align}
\mathcal{K}_m = \textrm{span}\left\{\ket{\psi(0)}, \hat{H}\ket{\psi(0)}, \hat{H}^2\ket{\psi(0)}, \dots, \right. \nonumber \\
\left. \hat{H}^{m-1}\ket{\psi(0)}\right\}.
\end{align} 
The optimal approximation is obtained by an Arnoldi decomposition procedure of the upper Hessenberg matrix $A_m$, defined as $A_m \defeq V^T_mHV_m$, where $V_m$ corresponds to the orthonormal basis resulting from the decomposition. $A_m$ can be seen as the projection of $\hat{H}$ onto $\mathcal{K}_m$ with respect to the basis $V_m$. In the previous description $m$ is the dimension of the Krylov subspace. In principle, the Arnoldi decomposition procedure can be replaced by a three-term Lanczos recursion for the specific case of Hermitian matrices. The latter amounts to a more efficient algorithm, yet to one that may suffer from numerical instabilities for ill-conditioned matrices.

The desired solution is then approximated by
\begin{equation}
\ket{\psi(t)}\approx V_{m}\exp(-\textrm{i}tA_m)\ket{e_1},
\end{equation}
where $\ket{e_1}$ is the first unit vector of the Krylov subspace. The approximation becomes an exact solution when $m \geq \mathcal{D}$, however, the method has been proven to be accurate even if $m \ll \mathcal{D}$ for short enough time-steps~\cite{expokit,brenes2019massively}. For the particular case when $m \ll \mathcal{D}$, the much smaller matrix exponential $\exp(-\textrm{i}tA_m)$ can be evaluated using standard numerical techniques, such as a Pad\`e approximation with a scaling-and-squaring algorithm. The error in the method behaves like $\mathcal{O}(e^{m-t||A||_{2}}(t||A||_{2}/m)^m)$ when $m \leq 2t||A||_{2}$, which indicates that the technique can be applied successfully if a time-stepping strategy is implemented along with error estimations~\cite{expokiterror}. In practice, the dimension of the Krylov subspace $m$ is a free parameter of the simulation, while the time-step is estimated such that the above {\em a priori} error estimation is kept under control.  

We finalise this discussion by remarking that for the specific case of $c(t)$, evaluating terms of the form $\braket{\psi | [\hat{O}(t), \hat{O}]^2 | \psi}$ is more complicated, since it requires running both forwards and backwards time evolution of
operators $\hat{O}$ acting on pure states $\ket{\psi}$, where the $\ket{\psi}$ are typical states introduced in Sec.~\ref{sec:scaling}. Yet, this procedure can be carried out efficiently dividing the entire time evolution into several time-steps~\cite{Luitz:2017}. 

\bibliography{bibliography.bib}

\begin{thebibliography}{60}%
\makeatletter
\providecommand \@ifxundefined [1]{%
 \@ifx{#1\undefined}
}%
\providecommand \@ifnum [1]{%
 \ifnum #1\expandafter \@firstoftwo
 \else \expandafter \@secondoftwo
 \fi
}%
\providecommand \@ifx [1]{%
 \ifx #1\expandafter \@firstoftwo
 \else \expandafter \@secondoftwo
 \fi
}%
\providecommand \natexlab [1]{#1}%
\providecommand \enquote  [1]{``#1''}%
\providecommand \bibnamefont  [1]{#1}%
\providecommand \bibfnamefont [1]{#1}%
\providecommand \citenamefont [1]{#1}%
\providecommand \href@noop [0]{\@secondoftwo}%
\providecommand \href [0]{\begingroup \@sanitize@url \@href}%
\providecommand \@href[1]{\@@startlink{#1}\@@href}%
\providecommand \@@href[1]{\endgroup#1\@@endlink}%
\providecommand \@sanitize@url [0]{\catcode `\\12\catcode `\$12\catcode
  `\&12\catcode `\#12\catcode `\^12\catcode `\_12\catcode `\%12\relax}%
\providecommand \@@startlink[1]{}%
\providecommand \@@endlink[0]{}%
\providecommand \url  [0]{\begingroup\@sanitize@url \@url }%
\providecommand \@url [1]{\endgroup\@href {#1}{\urlprefix }}%
\providecommand \urlprefix  [0]{URL }%
\providecommand \Eprint [0]{\href }%
\providecommand \doibase [0]{http://dx.doi.org/}%
\providecommand \selectlanguage [0]{\@gobble}%
\providecommand \bibinfo  [0]{\@secondoftwo}%
\providecommand \bibfield  [0]{\@secondoftwo}%
\providecommand \translation [1]{[#1]}%
\providecommand \BibitemOpen [0]{}%
\providecommand \bibitemStop [0]{}%
\providecommand \bibitemNoStop [0]{.\EOS\space}%
\providecommand \EOS [0]{\spacefactor3000\relax}%
\providecommand \BibitemShut  [1]{\csname bibitem#1\endcsname}%
\let\auto@bib@innerbib\@empty
%</preamble>
\bibitem [{\citenamefont {Schr{\"o}dinger}(1927)}]{Schrodinger:1927}%
  \BibitemOpen
  \bibfield  {author} {\bibinfo {author} {\bibfnamefont {E.}~\bibnamefont
  {Schr{\"o}dinger}},\ }\href {\doibase 10.1002/andp.19273881504} {\bibfield
  {journal} {\bibinfo  {journal} {Annalen der Physik}\ }\textbf {\bibinfo
  {volume} {388}},\ \bibinfo {pages} {956} (\bibinfo {year}
  {1927})}\BibitemShut {NoStop}%
\bibitem [{\citenamefont {Neumann}(1929)}]{Vonneumann:1929}%
  \BibitemOpen
  \bibfield  {author} {\bibinfo {author} {\bibfnamefont {J.~v.}\ \bibnamefont
  {Neumann}},\ }\href {\doibase 10.1007/BF01339852} {\bibfield  {journal}
  {\bibinfo  {journal} {Zeitschrift f{\"u}r Physik}\ }\textbf {\bibinfo
  {volume} {57}},\ \bibinfo {pages} {30} (\bibinfo {year} {1929})}\BibitemShut
  {NoStop}%
\bibitem [{\citenamefont {Rigol}\ \emph {et~al.}(2008)\citenamefont {Rigol},
  \citenamefont {Dunjko},\ and\ \citenamefont {Olshanii}}]{Rigol:2008}%
  \BibitemOpen
  \bibfield  {author} {\bibinfo {author} {\bibfnamefont {M.}~\bibnamefont
  {Rigol}}, \bibinfo {author} {\bibfnamefont {V.}~\bibnamefont {Dunjko}}, \
  and\ \bibinfo {author} {\bibfnamefont {M.}~\bibnamefont {Olshanii}},\ }\href
  {https://doi.org/10.1038/nature06838} {\bibfield  {journal} {\bibinfo
  {journal} {Nature}\ }\textbf {\bibinfo {volume} {452}},\ \bibinfo {pages}
  {854} (\bibinfo {year} {2008})}\BibitemShut {NoStop}%
\bibitem [{\citenamefont {Polkovnikov}\ \emph {et~al.}(2011)\citenamefont
  {Polkovnikov}, \citenamefont {Sengupta}, \citenamefont {Silva},\ and\
  \citenamefont {Vengalattore}}]{Polkovnikov:2011}%
  \BibitemOpen
  \bibfield  {author} {\bibinfo {author} {\bibfnamefont {A.}~\bibnamefont
  {Polkovnikov}}, \bibinfo {author} {\bibfnamefont {K.}~\bibnamefont
  {Sengupta}}, \bibinfo {author} {\bibfnamefont {A.}~\bibnamefont {Silva}}, \
  and\ \bibinfo {author} {\bibfnamefont {M.}~\bibnamefont {Vengalattore}},\
  }\href {https://doi.org/10.1103/RevModPhys.83.863} {\bibfield  {journal}
  {\bibinfo  {journal} {Reviews of Modern Physics}\ }\textbf {\bibinfo {volume}
  {83}},\ \bibinfo {pages} {863} (\bibinfo {year} {2011})}\BibitemShut
  {NoStop}%
\bibitem [{\citenamefont {Eisert}\ \emph {et~al.}(2015)\citenamefont {Eisert},
  \citenamefont {Friesdorf},\ and\ \citenamefont
  {Gogolin}}]{eisert2015quantum}%
  \BibitemOpen
  \bibfield  {author} {\bibinfo {author} {\bibfnamefont {J.}~\bibnamefont
  {Eisert}}, \bibinfo {author} {\bibfnamefont {M.}~\bibnamefont {Friesdorf}}, \
  and\ \bibinfo {author} {\bibfnamefont {C.}~\bibnamefont {Gogolin}},\ }\href
  {https://www.nature.com/articles/nphys3215} {\bibfield  {journal} {\bibinfo
  {journal} {Nature Physics}\ }\textbf {\bibinfo {volume} {11}},\ \bibinfo
  {pages} {124} (\bibinfo {year} {2015})}\BibitemShut {NoStop}%
\bibitem [{\citenamefont {Borgonovi}\ \emph {et~al.}(2016)\citenamefont
  {Borgonovi}, \citenamefont {Izrailev}, \citenamefont {Santos},\ and\
  \citenamefont {Zelevinsky}}]{Borgonovi:2016}%
  \BibitemOpen
  \bibfield  {author} {\bibinfo {author} {\bibfnamefont {F.}~\bibnamefont
  {Borgonovi}}, \bibinfo {author} {\bibfnamefont {F.}~\bibnamefont {Izrailev}},
  \bibinfo {author} {\bibfnamefont {L.}~\bibnamefont {Santos}}, \ and\ \bibinfo
  {author} {\bibfnamefont {V.}~\bibnamefont {Zelevinsky}},\ }\href
  {http://dx.doi.org/10.1016/j.physrep.2016.02.005} {\bibfield  {journal}
  {\bibinfo  {journal} {Physics Reports}\ }\textbf {\bibinfo {volume} {626}},\
  \bibinfo {pages} {1} (\bibinfo {year} {2016})}\BibitemShut {NoStop}%
\bibitem [{\citenamefont {D'Alessio}\ \emph {et~al.}(2016)\citenamefont
  {D'Alessio}, \citenamefont {Kafri}, \citenamefont {Polkovnikov},\ and\
  \citenamefont {Rigol}}]{Alessio:2016}%
  \BibitemOpen
  \bibfield  {author} {\bibinfo {author} {\bibfnamefont {L.}~\bibnamefont
  {D'Alessio}}, \bibinfo {author} {\bibfnamefont {Y.}~\bibnamefont {Kafri}},
  \bibinfo {author} {\bibfnamefont {A.}~\bibnamefont {Polkovnikov}}, \ and\
  \bibinfo {author} {\bibfnamefont {M.}~\bibnamefont {Rigol}},\ }\href
  {\doibase 10.1080/00018732.2016.1198134} {\bibfield  {journal} {\bibinfo
  {journal} {Advances in Physics}\ }\textbf {\bibinfo {volume} {65}},\ \bibinfo
  {pages} {239} (\bibinfo {year} {2016})}\BibitemShut {NoStop}%
\bibitem [{\citenamefont {Kinoshita}\ \emph {et~al.}(2006)\citenamefont
  {Kinoshita}, \citenamefont {Wenger},\ and\ \citenamefont
  {Weiss}}]{Kinoshita:2006}%
  \BibitemOpen
  \bibfield  {author} {\bibinfo {author} {\bibfnamefont {T.}~\bibnamefont
  {Kinoshita}}, \bibinfo {author} {\bibfnamefont {T.}~\bibnamefont {Wenger}}, \
  and\ \bibinfo {author} {\bibfnamefont {D.~S.}\ \bibnamefont {Weiss}},\ }\href
  {https://doi.org/10.1038/nature04693} {\bibfield  {journal} {\bibinfo
  {journal} {Nature}\ }\textbf {\bibinfo {volume} {440}},\ \bibinfo {pages}
  {900} (\bibinfo {year} {2006})}\BibitemShut {NoStop}%
\bibitem [{\citenamefont {Lewenstein}\ \emph {et~al.}(2007)\citenamefont
  {Lewenstein}, \citenamefont {Sanpera}, \citenamefont {Ahufinger},
  \citenamefont {Damski}, \citenamefont {Sen},\ and\ \citenamefont
  {Sen}}]{Lewenstein:2007}%
  \BibitemOpen
  \bibfield  {author} {\bibinfo {author} {\bibfnamefont {M.}~\bibnamefont
  {Lewenstein}}, \bibinfo {author} {\bibfnamefont {A.}~\bibnamefont {Sanpera}},
  \bibinfo {author} {\bibfnamefont {V.}~\bibnamefont {Ahufinger}}, \bibinfo
  {author} {\bibfnamefont {B.}~\bibnamefont {Damski}}, \bibinfo {author}
  {\bibfnamefont {A.}~\bibnamefont {Sen}}, \ and\ \bibinfo {author}
  {\bibfnamefont {U.}~\bibnamefont {Sen}},\ }\href
  {http://dx.doi.org/10.1080/00018730701223200} {\bibfield  {journal} {\bibinfo
   {journal} {Advances in Physics}\ }\textbf {\bibinfo {volume} {56}},\
  \bibinfo {pages} {243} (\bibinfo {year} {2007})}\BibitemShut {NoStop}%
\bibitem [{\citenamefont {Bloch}\ \emph {et~al.}(2012)\citenamefont {Bloch},
  \citenamefont {Dalibard},\ and\ \citenamefont {Nascimbene}}]{Bloch:2012}%
  \BibitemOpen
  \bibfield  {author} {\bibinfo {author} {\bibfnamefont {I.}~\bibnamefont
  {Bloch}}, \bibinfo {author} {\bibfnamefont {J.}~\bibnamefont {Dalibard}}, \
  and\ \bibinfo {author} {\bibfnamefont {S.}~\bibnamefont {Nascimbene}},\
  }\href {https://doi.org/10.1038/nphys2259} {\bibfield  {journal} {\bibinfo
  {journal} {Nature Physics}\ }\textbf {\bibinfo {volume} {8}},\ \bibinfo
  {pages} {267} (\bibinfo {year} {2012})}\BibitemShut {NoStop}%
\bibitem [{\citenamefont {Lebowitz}\ and\ \citenamefont
  {Penrose}(1973)}]{lebowitz1973modern}%
  \BibitemOpen
  \bibfield  {author} {\bibinfo {author} {\bibfnamefont {J.~L.}\ \bibnamefont
  {Lebowitz}}\ and\ \bibinfo {author} {\bibfnamefont {O.}~\bibnamefont
  {Penrose}},\ }\href
  {https://physicstoday.scitation.org/doi/10.1063/1.3127948} {\bibfield
  {journal} {\bibinfo  {journal} {Physics Today}\ }\textbf {\bibinfo {volume}
  {26}},\ \bibinfo {pages} {23} (\bibinfo {year} {1973})}\BibitemShut {NoStop}%
\bibitem [{\citenamefont {Deutsch}(1991)}]{Deutsch:1991}%
  \BibitemOpen
  \bibfield  {author} {\bibinfo {author} {\bibfnamefont {J.~M.}\ \bibnamefont
  {Deutsch}},\ }\href {\doibase 10.1103/PhysRevA.43.2046} {\bibfield  {journal}
  {\bibinfo  {journal} {Physical Review A}\ }\textbf {\bibinfo {volume} {43}},\
  \bibinfo {pages} {2046} (\bibinfo {year} {1991})}\BibitemShut {NoStop}%
\bibitem [{\citenamefont {Srednicki}(1994)}]{Srednicki:1994}%
  \BibitemOpen
  \bibfield  {author} {\bibinfo {author} {\bibfnamefont {M.}~\bibnamefont
  {Srednicki}},\ }\href {\doibase 10.1103/PhysRevE.50.888} {\bibfield
  {journal} {\bibinfo  {journal} {Physical Review E}\ }\textbf {\bibinfo
  {volume} {50}},\ \bibinfo {pages} {888} (\bibinfo {year} {1994})}\BibitemShut
  {NoStop}%
\bibitem [{\citenamefont {Srednicki}(1999)}]{Srednicki:1999}%
  \BibitemOpen
  \bibfield  {author} {\bibinfo {author} {\bibfnamefont {M.}~\bibnamefont
  {Srednicki}},\ }\href {http://dx.doi.org/10.1088/0305-4470/32/7/007}
  {\bibfield  {journal} {\bibinfo  {journal} {Journal of Physics A:
  Mathematical and General}\ }\textbf {\bibinfo {volume} {32}},\ \bibinfo
  {pages} {1163} (\bibinfo {year} {1999})}\BibitemShut {NoStop}%
\bibitem [{\citenamefont {Swingle}(2018)}]{swingle2018}%
  \BibitemOpen
  \bibfield  {author} {\bibinfo {author} {\bibfnamefont {B.}~\bibnamefont
  {Swingle}},\ }\href {https://www.nature.com/articles/s41567-018-0295-5}
  {\bibfield  {journal} {\bibinfo  {journal} {Nature Physics}\ }\textbf
  {\bibinfo {volume} {14}},\ \bibinfo {pages} {988} (\bibinfo {year}
  {2018})}\BibitemShut {NoStop}%
\bibitem [{\citenamefont {Hosur}\ \emph {et~al.}(2016)\citenamefont {Hosur},
  \citenamefont {Qi}, \citenamefont {Roberts},\ and\ \citenamefont
  {Yoshida}}]{Hosur2016Chaos}%
  \BibitemOpen
  \bibfield  {author} {\bibinfo {author} {\bibfnamefont {P.}~\bibnamefont
  {Hosur}}, \bibinfo {author} {\bibfnamefont {X.-L.}\ \bibnamefont {Qi}},
  \bibinfo {author} {\bibfnamefont {D.~A.}\ \bibnamefont {Roberts}}, \ and\
  \bibinfo {author} {\bibfnamefont {B.}~\bibnamefont {Yoshida}},\ }\href
  {\doibase 10.1007/JHEP02(2016)004} {\bibfield  {journal} {\bibinfo  {journal}
  {Journal of High Energy Physics}\ }\textbf {\bibinfo {volume} {2016}},\
  \bibinfo {pages} {4} (\bibinfo {year} {2016})}\BibitemShut {NoStop}%
\bibitem [{\citenamefont {Parker}\ \emph {et~al.}(2019)\citenamefont {Parker},
  \citenamefont {Cao}, \citenamefont {Avdoshkin}, \citenamefont {Scaffidi},\
  and\ \citenamefont {Altman}}]{Parker19}%
  \BibitemOpen
  \bibfield  {author} {\bibinfo {author} {\bibfnamefont {D.~E.}\ \bibnamefont
  {Parker}}, \bibinfo {author} {\bibfnamefont {X.}~\bibnamefont {Cao}},
  \bibinfo {author} {\bibfnamefont {A.}~\bibnamefont {Avdoshkin}}, \bibinfo
  {author} {\bibfnamefont {T.}~\bibnamefont {Scaffidi}}, \ and\ \bibinfo
  {author} {\bibfnamefont {E.}~\bibnamefont {Altman}},\ }\href {\doibase
  10.1103/PhysRevX.9.041017} {\bibfield  {journal} {\bibinfo  {journal}
  {Physical Review X}\ }\textbf {\bibinfo {volume} {9}},\ \bibinfo {pages}
  {041017} (\bibinfo {year} {2019})}\BibitemShut {NoStop}%
\bibitem [{\citenamefont {Larkin}\ and\ \citenamefont
  {Ovchinnikov}(1969)}]{larkin1969quasiclassical}%
  \BibitemOpen
  \bibfield  {author} {\bibinfo {author} {\bibfnamefont {A.}~\bibnamefont
  {Larkin}}\ and\ \bibinfo {author} {\bibfnamefont {Y.~N.}\ \bibnamefont
  {Ovchinnikov}},\ }\href
  {http://www.jetp.ac.ru/cgi-bin/e/index/e/28/6/p1200?a=list} {\bibfield
  {journal} {\bibinfo  {journal} {Soviet Physics JETP}\ }\textbf {\bibinfo
  {volume} {28}},\ \bibinfo {pages} {1200} (\bibinfo {year}
  {1969})}\BibitemShut {NoStop}%
\bibitem [{\citenamefont {Shenker}\ and\ \citenamefont
  {Stanford}(2014)}]{shenker2014black}%
  \BibitemOpen
  \bibfield  {author} {\bibinfo {author} {\bibfnamefont {S.~H.}\ \bibnamefont
  {Shenker}}\ and\ \bibinfo {author} {\bibfnamefont {D.}~\bibnamefont
  {Stanford}},\ }\href
  {https://link.springer.com/article/10.1007%2FJHEP03%282014%29067} {\bibfield
  {journal} {\bibinfo  {journal} {Journal of High Energy Physics}\ }\textbf
  {\bibinfo {volume} {2014}},\ \bibinfo {pages} {67} (\bibinfo {year}
  {2014})}\BibitemShut {NoStop}%
\bibitem [{\citenamefont {Maldacena}\ \emph {et~al.}(2016)\citenamefont
  {Maldacena}, \citenamefont {Shenker},\ and\ \citenamefont
  {Stanford}}]{maldacena2016}%
  \BibitemOpen
  \bibfield  {author} {\bibinfo {author} {\bibfnamefont {J.}~\bibnamefont
  {Maldacena}}, \bibinfo {author} {\bibfnamefont {S.~H.}\ \bibnamefont
  {Shenker}}, \ and\ \bibinfo {author} {\bibfnamefont {D.}~\bibnamefont
  {Stanford}},\ }\href
  {https://link.springer.com/article/10.1007%2FJHEP08%282016%29106} {\bibfield
  {journal} {\bibinfo  {journal} {Journal of High Energy Physics}\ }\textbf
  {\bibinfo {volume} {2016}},\ \bibinfo {pages} {106} (\bibinfo {year}
  {2016})}\BibitemShut {NoStop}%
\bibitem [{\citenamefont {G{\"a}rttner}\ \emph {et~al.}(2017)\citenamefont
  {G{\"a}rttner}, \citenamefont {Bohnet}, \citenamefont {Safavi-Naini},
  \citenamefont {Wall}, \citenamefont {Bollinger},\ and\ \citenamefont
  {Rey}}]{garttner2017}%
  \BibitemOpen
  \bibfield  {author} {\bibinfo {author} {\bibfnamefont {M.}~\bibnamefont
  {G{\"a}rttner}}, \bibinfo {author} {\bibfnamefont {J.~G.}\ \bibnamefont
  {Bohnet}}, \bibinfo {author} {\bibfnamefont {A.}~\bibnamefont
  {Safavi-Naini}}, \bibinfo {author} {\bibfnamefont {M.~L.}\ \bibnamefont
  {Wall}}, \bibinfo {author} {\bibfnamefont {J.~J.}\ \bibnamefont {Bollinger}},
  \ and\ \bibinfo {author} {\bibfnamefont {A.~M.}\ \bibnamefont {Rey}},\ }\href
  {https://www.nature.com/articles/nphys4119} {\bibfield  {journal} {\bibinfo
  {journal} {Nature Physics}\ }\textbf {\bibinfo {volume} {13}},\ \bibinfo
  {pages} {781} (\bibinfo {year} {2017})}\BibitemShut {NoStop}%
\bibitem [{\citenamefont {Li}\ \emph {et~al.}(2017)\citenamefont {Li},
  \citenamefont {Fan}, \citenamefont {Wang}, \citenamefont {Ye}, \citenamefont
  {Zeng}, \citenamefont {Zhai}, \citenamefont {Peng},\ and\ \citenamefont
  {Du}}]{li2020}%
  \BibitemOpen
  \bibfield  {author} {\bibinfo {author} {\bibfnamefont {J.}~\bibnamefont
  {Li}}, \bibinfo {author} {\bibfnamefont {R.}~\bibnamefont {Fan}}, \bibinfo
  {author} {\bibfnamefont {H.}~\bibnamefont {Wang}}, \bibinfo {author}
  {\bibfnamefont {B.}~\bibnamefont {Ye}}, \bibinfo {author} {\bibfnamefont
  {B.}~\bibnamefont {Zeng}}, \bibinfo {author} {\bibfnamefont {H.}~\bibnamefont
  {Zhai}}, \bibinfo {author} {\bibfnamefont {X.}~\bibnamefont {Peng}}, \ and\
  \bibinfo {author} {\bibfnamefont {J.}~\bibnamefont {Du}},\ }\href {\doibase
  10.1103/PhysRevX.7.031011} {\bibfield  {journal} {\bibinfo  {journal}
  {Physical Review X}\ }\textbf {\bibinfo {volume} {7}},\ \bibinfo {pages}
  {031011} (\bibinfo {year} {2017})}\BibitemShut {NoStop}%
\bibitem [{\citenamefont {Nie}\ \emph {et~al.}(2020)\citenamefont {Nie},
  \citenamefont {Wei}, \citenamefont {Chen}, \citenamefont {Zhang},
  \citenamefont {Zhao}, \citenamefont {Qiu}, \citenamefont {Tian},
  \citenamefont {Ji}, \citenamefont {Xin}, \citenamefont {Lu},\ and\
  \citenamefont {Li}}]{Nie20}%
  \BibitemOpen
  \bibfield  {author} {\bibinfo {author} {\bibfnamefont {X.}~\bibnamefont
  {Nie}}, \bibinfo {author} {\bibfnamefont {B.-B.}\ \bibnamefont {Wei}},
  \bibinfo {author} {\bibfnamefont {X.}~\bibnamefont {Chen}}, \bibinfo {author}
  {\bibfnamefont {Z.}~\bibnamefont {Zhang}}, \bibinfo {author} {\bibfnamefont
  {X.}~\bibnamefont {Zhao}}, \bibinfo {author} {\bibfnamefont {C.}~\bibnamefont
  {Qiu}}, \bibinfo {author} {\bibfnamefont {Y.}~\bibnamefont {Tian}}, \bibinfo
  {author} {\bibfnamefont {Y.}~\bibnamefont {Ji}}, \bibinfo {author}
  {\bibfnamefont {T.}~\bibnamefont {Xin}}, \bibinfo {author} {\bibfnamefont
  {D.}~\bibnamefont {Lu}}, \ and\ \bibinfo {author} {\bibfnamefont
  {J.}~\bibnamefont {Li}},\ }\href {\doibase 10.1103/PhysRevLett.124.250601}
  {\bibfield  {journal} {\bibinfo  {journal} {Physical Review Letters}\
  }\textbf {\bibinfo {volume} {124}},\ \bibinfo {pages} {250601} (\bibinfo
  {year} {2020})}\BibitemShut {NoStop}%
\bibitem [{\citenamefont {Mi}\ \emph {et~al.}(2021)\citenamefont {Mi},
  \citenamefont {Roushan}, \citenamefont {Quintana}, \citenamefont {Mandra},
  \citenamefont {Marshall}, \citenamefont {Neill}, \citenamefont {Arute},
  \citenamefont {Arya}, \citenamefont {Atalaya}, \citenamefont {Babbush} \emph
  {et~al.}}]{mi2021}%
  \BibitemOpen
  \bibfield  {author} {\bibinfo {author} {\bibfnamefont {X.}~\bibnamefont
  {Mi}}, \bibinfo {author} {\bibfnamefont {P.}~\bibnamefont {Roushan}},
  \bibinfo {author} {\bibfnamefont {C.}~\bibnamefont {Quintana}}, \bibinfo
  {author} {\bibfnamefont {S.}~\bibnamefont {Mandra}}, \bibinfo {author}
  {\bibfnamefont {J.}~\bibnamefont {Marshall}}, \bibinfo {author}
  {\bibfnamefont {C.}~\bibnamefont {Neill}}, \bibinfo {author} {\bibfnamefont
  {F.}~\bibnamefont {Arute}}, \bibinfo {author} {\bibfnamefont
  {K.}~\bibnamefont {Arya}}, \bibinfo {author} {\bibfnamefont {J.}~\bibnamefont
  {Atalaya}}, \bibinfo {author} {\bibfnamefont {R.}~\bibnamefont {Babbush}},
  \emph {et~al.},\ }\href {https://arxiv.org/abs/2101.08870} {\bibfield
  {journal} {\bibinfo  {journal} {arXiv preprint arXiv:2101.08870}\ } (\bibinfo
  {year} {2021})}\BibitemShut {NoStop}%
\bibitem [{\citenamefont {Foini}\ and\ \citenamefont
  {Kurchan}(2019{\natexlab{a}})}]{Foini2019}%
  \BibitemOpen
  \bibfield  {author} {\bibinfo {author} {\bibfnamefont {L.}~\bibnamefont
  {Foini}}\ and\ \bibinfo {author} {\bibfnamefont {J.}~\bibnamefont
  {Kurchan}},\ }\href {\doibase 10.1103/PhysRevE.99.042139} {\bibfield
  {journal} {\bibinfo  {journal} {Physical Review E}\ }\textbf {\bibinfo
  {volume} {99}},\ \bibinfo {pages} {042139} (\bibinfo {year}
  {2019}{\natexlab{a}})}\BibitemShut {NoStop}%
\bibitem [{\citenamefont {Murthy}\ and\ \citenamefont
  {Srednicki}(2019)}]{Murthy19}%
  \BibitemOpen
  \bibfield  {author} {\bibinfo {author} {\bibfnamefont {C.}~\bibnamefont
  {Murthy}}\ and\ \bibinfo {author} {\bibfnamefont {M.}~\bibnamefont
  {Srednicki}},\ }\href {\doibase 10.1103/PhysRevLett.123.230606} {\bibfield
  {journal} {\bibinfo  {journal} {Physical Review Letters}\ }\textbf {\bibinfo
  {volume} {123}},\ \bibinfo {pages} {230606} (\bibinfo {year}
  {2019})}\BibitemShut {NoStop}%
\bibitem [{\citenamefont {Chan}\ \emph {et~al.}(2019)\citenamefont {Chan},
  \citenamefont {De~Luca},\ and\ \citenamefont {Chalker}}]{Chan2019}%
  \BibitemOpen
  \bibfield  {author} {\bibinfo {author} {\bibfnamefont {A.}~\bibnamefont
  {Chan}}, \bibinfo {author} {\bibfnamefont {A.}~\bibnamefont {De~Luca}}, \
  and\ \bibinfo {author} {\bibfnamefont {J.~T.}\ \bibnamefont {Chalker}},\
  }\href {\doibase 10.1103/PhysRevLett.122.220601} {\bibfield  {journal}
  {\bibinfo  {journal} {Physical Review Letters}\ }\textbf {\bibinfo {volume}
  {122}},\ \bibinfo {pages} {220601} (\bibinfo {year} {2019})}\BibitemShut
  {NoStop}%
\bibitem [{\citenamefont {Richter}\ \emph {et~al.}(2020)\citenamefont
  {Richter}, \citenamefont {Dymarsky}, \citenamefont {Steinigeweg},\ and\
  \citenamefont {Gemmer}}]{Richter2020}%
  \BibitemOpen
  \bibfield  {author} {\bibinfo {author} {\bibfnamefont {J.}~\bibnamefont
  {Richter}}, \bibinfo {author} {\bibfnamefont {A.}~\bibnamefont {Dymarsky}},
  \bibinfo {author} {\bibfnamefont {R.}~\bibnamefont {Steinigeweg}}, \ and\
  \bibinfo {author} {\bibfnamefont {J.}~\bibnamefont {Gemmer}},\ }\href
  {\doibase 10.1103/PhysRevE.102.042127} {\bibfield  {journal} {\bibinfo
  {journal} {Physical Review E}\ }\textbf {\bibinfo {volume} {102}},\ \bibinfo
  {pages} {042127} (\bibinfo {year} {2020})}\BibitemShut {NoStop}%
\bibitem [{\citenamefont {Khatami}\ \emph {et~al.}(2013)\citenamefont
  {Khatami}, \citenamefont {Pupillo}, \citenamefont {Srednicki},\ and\
  \citenamefont {Rigol}}]{Khatami2013}%
  \BibitemOpen
  \bibfield  {author} {\bibinfo {author} {\bibfnamefont {E.}~\bibnamefont
  {Khatami}}, \bibinfo {author} {\bibfnamefont {G.}~\bibnamefont {Pupillo}},
  \bibinfo {author} {\bibfnamefont {M.}~\bibnamefont {Srednicki}}, \ and\
  \bibinfo {author} {\bibfnamefont {M.}~\bibnamefont {Rigol}},\ }\href
  {\doibase 10.1103/PhysRevLett.111.050403} {\bibfield  {journal} {\bibinfo
  {journal} {Physical Review Letters}\ }\textbf {\bibinfo {volume} {111}},\
  \bibinfo {pages} {050403} (\bibinfo {year} {2013})}\BibitemShut {NoStop}%
\bibitem [{\citenamefont {Kim}\ and\ \citenamefont {Huse}(2013)}]{Kim2013}%
  \BibitemOpen
  \bibfield  {author} {\bibinfo {author} {\bibfnamefont {H.}~\bibnamefont
  {Kim}}\ and\ \bibinfo {author} {\bibfnamefont {D.~A.}\ \bibnamefont {Huse}},\
  }\href {\doibase 10.1103/PhysRevLett.111.127205} {\bibfield  {journal}
  {\bibinfo  {journal} {Physical Review Letters}\ }\textbf {\bibinfo {volume}
  {111}},\ \bibinfo {pages} {127205} (\bibinfo {year} {2013})}\BibitemShut
  {NoStop}%
\bibitem [{\citenamefont {Beugeling}\ \emph {et~al.}(2015)\citenamefont
  {Beugeling}, \citenamefont {Moessner},\ and\ \citenamefont
  {Haque}}]{Beugeling2015}%
  \BibitemOpen
  \bibfield  {author} {\bibinfo {author} {\bibfnamefont {W.}~\bibnamefont
  {Beugeling}}, \bibinfo {author} {\bibfnamefont {R.}~\bibnamefont {Moessner}},
  \ and\ \bibinfo {author} {\bibfnamefont {M.}~\bibnamefont {Haque}},\ }\href
  {\doibase 10.1103/PhysRevE.91.012144} {\bibfield  {journal} {\bibinfo
  {journal} {Physical Review E}\ }\textbf {\bibinfo {volume} {91}},\ \bibinfo
  {pages} {012144} (\bibinfo {year} {2015})}\BibitemShut {NoStop}%
\bibitem [{\citenamefont {LeBlond}\ \emph {et~al.}(2019)\citenamefont
  {LeBlond}, \citenamefont {Mallayya}, \citenamefont {Vidmar},\ and\
  \citenamefont {Rigol}}]{Leblond:2019}%
  \BibitemOpen
  \bibfield  {author} {\bibinfo {author} {\bibfnamefont {T.}~\bibnamefont
  {LeBlond}}, \bibinfo {author} {\bibfnamefont {K.}~\bibnamefont {Mallayya}},
  \bibinfo {author} {\bibfnamefont {L.}~\bibnamefont {Vidmar}}, \ and\ \bibinfo
  {author} {\bibfnamefont {M.}~\bibnamefont {Rigol}},\ }\href {\doibase
  10.1103/PhysRevE.100.062134} {\bibfield  {journal} {\bibinfo  {journal}
  {Physical Review E}\ }\textbf {\bibinfo {volume} {100}},\ \bibinfo {pages}
  {062134} (\bibinfo {year} {2019})}\BibitemShut {NoStop}%
\bibitem [{\citenamefont {Khaymovich}\ \emph {et~al.}(2019)\citenamefont
  {Khaymovich}, \citenamefont {Haque},\ and\ \citenamefont
  {McClarty}}]{Khaymovich2019}%
  \BibitemOpen
  \bibfield  {author} {\bibinfo {author} {\bibfnamefont {I.~M.}\ \bibnamefont
  {Khaymovich}}, \bibinfo {author} {\bibfnamefont {M.}~\bibnamefont {Haque}}, \
  and\ \bibinfo {author} {\bibfnamefont {P.~A.}\ \bibnamefont {McClarty}},\
  }\href {\doibase 10.1103/PhysRevLett.122.070601} {\bibfield  {journal}
  {\bibinfo  {journal} {Physical Review Letters}\ }\textbf {\bibinfo {volume}
  {122}},\ \bibinfo {pages} {070601} (\bibinfo {year} {2019})}\BibitemShut
  {NoStop}%
\bibitem [{\citenamefont {Brenes}\ \emph
  {et~al.}(2020{\natexlab{a}})\citenamefont {Brenes}, \citenamefont {LeBlond},
  \citenamefont {Goold},\ and\ \citenamefont {Rigol}}]{BrenesLocal2020}%
  \BibitemOpen
  \bibfield  {author} {\bibinfo {author} {\bibfnamefont {M.}~\bibnamefont
  {Brenes}}, \bibinfo {author} {\bibfnamefont {T.}~\bibnamefont {LeBlond}},
  \bibinfo {author} {\bibfnamefont {J.}~\bibnamefont {Goold}}, \ and\ \bibinfo
  {author} {\bibfnamefont {M.}~\bibnamefont {Rigol}},\ }\href {\doibase
  10.1103/PhysRevLett.125.070605} {\bibfield  {journal} {\bibinfo  {journal}
  {Physical Review Letters}\ }\textbf {\bibinfo {volume} {125}},\ \bibinfo
  {pages} {070605} (\bibinfo {year} {2020}{\natexlab{a}})}\BibitemShut
  {NoStop}%
\bibitem [{\citenamefont {LeBlond}\ and\ \citenamefont
  {Rigol}(2020)}]{Leblond:2020}%
  \BibitemOpen
  \bibfield  {author} {\bibinfo {author} {\bibfnamefont {T.}~\bibnamefont
  {LeBlond}}\ and\ \bibinfo {author} {\bibfnamefont {M.}~\bibnamefont
  {Rigol}},\ }\href {\doibase 10.1103/PhysRevE.102.062113} {\bibfield
  {journal} {\bibinfo  {journal} {Physical Review E}\ }\textbf {\bibinfo
  {volume} {102}},\ \bibinfo {pages} {062113} (\bibinfo {year}
  {2020})}\BibitemShut {NoStop}%
\bibitem [{\citenamefont {Santos}\ \emph {et~al.}(2020)\citenamefont {Santos},
  \citenamefont {P\'erez-Bernal},\ and\ \citenamefont
  {Torres-Herrera}}]{Santos2020}%
  \BibitemOpen
  \bibfield  {author} {\bibinfo {author} {\bibfnamefont {L.~F.}\ \bibnamefont
  {Santos}}, \bibinfo {author} {\bibfnamefont {F.}~\bibnamefont
  {P\'erez-Bernal}}, \ and\ \bibinfo {author} {\bibfnamefont {E.~J.}\
  \bibnamefont {Torres-Herrera}},\ }\href {\doibase
  10.1103/PhysRevResearch.2.043034} {\bibfield  {journal} {\bibinfo  {journal}
  {Physical Review Research}\ }\textbf {\bibinfo {volume} {2}},\ \bibinfo
  {pages} {043034} (\bibinfo {year} {2020})}\BibitemShut {NoStop}%
\bibitem [{\citenamefont {Brenes}\ \emph
  {et~al.}(2020{\natexlab{b}})\citenamefont {Brenes}, \citenamefont {Goold},\
  and\ \citenamefont {Rigol}}]{BrenesFreq2020}%
  \BibitemOpen
  \bibfield  {author} {\bibinfo {author} {\bibfnamefont {M.}~\bibnamefont
  {Brenes}}, \bibinfo {author} {\bibfnamefont {J.}~\bibnamefont {Goold}}, \
  and\ \bibinfo {author} {\bibfnamefont {M.}~\bibnamefont {Rigol}},\ }\href
  {\doibase 10.1103/PhysRevB.102.075127} {\bibfield  {journal} {\bibinfo
  {journal} {Physical Review B}\ }\textbf {\bibinfo {volume} {102}},\ \bibinfo
  {pages} {075127} (\bibinfo {year} {2020}{\natexlab{b}})}\BibitemShut
  {NoStop}%
\bibitem [{\citenamefont {Luitz}\ and\ \citenamefont
  {Bar~Lev}(2016)}]{Luitz2016Anomalous}%
  \BibitemOpen
  \bibfield  {author} {\bibinfo {author} {\bibfnamefont {D.~J.}\ \bibnamefont
  {Luitz}}\ and\ \bibinfo {author} {\bibfnamefont {Y.}~\bibnamefont
  {Bar~Lev}},\ }\href {\doibase 10.1103/PhysRevLett.117.170404} {\bibfield
  {journal} {\bibinfo  {journal} {Physical Review Letters}\ }\textbf {\bibinfo
  {volume} {117}},\ \bibinfo {pages} {170404} (\bibinfo {year}
  {2016})}\BibitemShut {NoStop}%
\bibitem [{\citenamefont {Luitz}(2016)}]{Luitz2016Long}%
  \BibitemOpen
  \bibfield  {author} {\bibinfo {author} {\bibfnamefont {D.~J.}\ \bibnamefont
  {Luitz}},\ }\href {\doibase 10.1103/PhysRevB.93.134201} {\bibfield  {journal}
  {\bibinfo  {journal} {Phys. Rev. B}\ }\textbf {\bibinfo {volume} {93}},\
  \bibinfo {pages} {134201} (\bibinfo {year} {2016})}\BibitemShut {NoStop}%
\bibitem [{\citenamefont {Foini}\ and\ \citenamefont
  {Kurchan}(2019{\natexlab{b}})}]{Foini2019Eigenstate}%
  \BibitemOpen
  \bibfield  {author} {\bibinfo {author} {\bibfnamefont {L.}~\bibnamefont
  {Foini}}\ and\ \bibinfo {author} {\bibfnamefont {J.}~\bibnamefont
  {Kurchan}},\ }\href {\doibase 10.1103/PhysRevLett.123.260601} {\bibfield
  {journal} {\bibinfo  {journal} {Phys. Rev. Lett.}\ }\textbf {\bibinfo
  {volume} {123}},\ \bibinfo {pages} {260601} (\bibinfo {year}
  {2019}{\natexlab{b}})}\BibitemShut {NoStop}%
\bibitem [{\citenamefont {Mondaini}\ and\ \citenamefont
  {Rigol}(2017)}]{Mondaini2017}%
  \BibitemOpen
  \bibfield  {author} {\bibinfo {author} {\bibfnamefont {R.}~\bibnamefont
  {Mondaini}}\ and\ \bibinfo {author} {\bibfnamefont {M.}~\bibnamefont
  {Rigol}},\ }\href {\doibase 10.1103/PhysRevE.96.012157} {\bibfield  {journal}
  {\bibinfo  {journal} {Physical Review E}\ }\textbf {\bibinfo {volume} {96}},\
  \bibinfo {pages} {012157} (\bibinfo {year} {2017})}\BibitemShut {NoStop}%
\bibitem [{\citenamefont {Brenes}\ \emph
  {et~al.}(2020{\natexlab{c}})\citenamefont {Brenes}, \citenamefont
  {Pappalardi}, \citenamefont {Goold},\ and\ \citenamefont
  {Silva}}]{BrenesFisher2020}%
  \BibitemOpen
  \bibfield  {author} {\bibinfo {author} {\bibfnamefont {M.}~\bibnamefont
  {Brenes}}, \bibinfo {author} {\bibfnamefont {S.}~\bibnamefont {Pappalardi}},
  \bibinfo {author} {\bibfnamefont {J.}~\bibnamefont {Goold}}, \ and\ \bibinfo
  {author} {\bibfnamefont {A.}~\bibnamefont {Silva}},\ }\href {\doibase
  10.1103/PhysRevLett.124.040605} {\bibfield  {journal} {\bibinfo  {journal}
  {Physical Review Letters}\ }\textbf {\bibinfo {volume} {124}},\ \bibinfo
  {pages} {040605} (\bibinfo {year} {2020}{\natexlab{c}})}\BibitemShut
  {NoStop}%
\bibitem [{\citenamefont {Mehta}(2004)}]{mehta2004random}%
  \BibitemOpen
  \bibfield  {author} {\bibinfo {author} {\bibfnamefont {M.~L.}\ \bibnamefont
  {Mehta}},\ }\href
  {https://www.elsevier.com/books/random-matrices/lal-mehta/978-0-12-088409-4}
  {\emph {\bibinfo {title} {Random matrices}}}\ (\bibinfo  {publisher}
  {Elsevier},\ \bibinfo {year} {2004})\BibitemShut {NoStop}%
\bibitem [{\citenamefont {Livan}\ \emph {et~al.}(2018)\citenamefont {Livan},
  \citenamefont {Novaes},\ and\ \citenamefont {Vivo}}]{Livan2018introduction}%
  \BibitemOpen
  \bibfield  {author} {\bibinfo {author} {\bibfnamefont {G.}~\bibnamefont
  {Livan}}, \bibinfo {author} {\bibfnamefont {M.}~\bibnamefont {Novaes}}, \
  and\ \bibinfo {author} {\bibfnamefont {P.}~\bibnamefont {Vivo}},\ }\href
  {https://doi.org/10.1007/978-3-319-70885-0} {\emph {\bibinfo {title}
  {Introduction to Random Matrices}}}\ (\bibinfo  {publisher} {Springer
  International Publishing},\ \bibinfo {year} {2018})\BibitemShut {NoStop}%
\bibitem [{\citenamefont {Cotler}\ \emph {et~al.}(2017)\citenamefont {Cotler},
  \citenamefont {Hunter-Jones}, \citenamefont {Liu},\ and\ \citenamefont
  {Yoshida}}]{Cotler2017Chaos}%
  \BibitemOpen
  \bibfield  {author} {\bibinfo {author} {\bibfnamefont {J.}~\bibnamefont
  {Cotler}}, \bibinfo {author} {\bibfnamefont {N.}~\bibnamefont
  {Hunter-Jones}}, \bibinfo {author} {\bibfnamefont {J.}~\bibnamefont {Liu}}, \
  and\ \bibinfo {author} {\bibfnamefont {B.}~\bibnamefont {Yoshida}},\ }\href
  {\doibase 10.1007/JHEP11(2017)048} {\bibfield  {journal} {\bibinfo  {journal}
  {Journal of High Energy Physics}\ }\textbf {\bibinfo {volume} {2017}},\
  \bibinfo {pages} {48} (\bibinfo {year} {2017})}\BibitemShut {NoStop}%
\bibitem [{\citenamefont {Sachdev}(2011)}]{sachdev2011}%
  \BibitemOpen
  \bibfield  {author} {\bibinfo {author} {\bibfnamefont {S.}~\bibnamefont
  {Sachdev}},\ }\href
  {https://www.cambridge.org/core/books/quantum-phase-transitions/33C1C81500346005E54C1DE4223E5562}
  {\emph {\bibinfo {title} {Quantum phase transitions}}},\ \bibinfo {edition}
  {2nd}\ ed.\ (\bibinfo  {publisher} {Cambridge University Press},\ \bibinfo
  {address} {Cambridge},\ \bibinfo {year} {2011})\BibitemShut {NoStop}%
\bibitem [{\citenamefont {Goldstein}\ \emph {et~al.}(2006)\citenamefont
  {Goldstein}, \citenamefont {Lebowitz}, \citenamefont {Tumulka},\ and\
  \citenamefont {Zangh\`{\i}}}]{Goldstein:2006}%
  \BibitemOpen
  \bibfield  {author} {\bibinfo {author} {\bibfnamefont {S.}~\bibnamefont
  {Goldstein}}, \bibinfo {author} {\bibfnamefont {J.~L.}\ \bibnamefont
  {Lebowitz}}, \bibinfo {author} {\bibfnamefont {R.}~\bibnamefont {Tumulka}}, \
  and\ \bibinfo {author} {\bibfnamefont {N.}~\bibnamefont {Zangh\`{\i}}},\
  }\href {\doibase 10.1103/PhysRevLett.96.050403} {\bibfield  {journal}
  {\bibinfo  {journal} {Physical Review Letters}\ }\textbf {\bibinfo {volume}
  {96}},\ \bibinfo {pages} {050403} (\bibinfo {year} {2006})}\BibitemShut
  {NoStop}%
\bibitem [{\citenamefont {Popescu}\ \emph {et~al.}(2006)\citenamefont
  {Popescu}, \citenamefont {Short},\ and\ \citenamefont
  {Winter}}]{Popescu:2006}%
  \BibitemOpen
  \bibfield  {author} {\bibinfo {author} {\bibfnamefont {S.}~\bibnamefont
  {Popescu}}, \bibinfo {author} {\bibfnamefont {A.~J.}\ \bibnamefont {Short}},
  \ and\ \bibinfo {author} {\bibfnamefont {A.}~\bibnamefont {Winter}},\ }\href
  {\doibase 10.1038/nphys444} {\bibfield  {journal} {\bibinfo  {journal}
  {Nature Physics}\ }\textbf {\bibinfo {volume} {2}},\ \bibinfo {pages} {754}
  (\bibinfo {year} {2006})}\BibitemShut {NoStop}%
\bibitem [{\citenamefont {Luitz}\ and\ \citenamefont
  {Bar~Lev}(2017)}]{Luitz:2017}%
  \BibitemOpen
  \bibfield  {author} {\bibinfo {author} {\bibfnamefont {D.~J.}\ \bibnamefont
  {Luitz}}\ and\ \bibinfo {author} {\bibfnamefont {Y.}~\bibnamefont
  {Bar~Lev}},\ }\href {\doibase 10.1103/PhysRevB.96.020406} {\bibfield
  {journal} {\bibinfo  {journal} {Physical Review B}\ }\textbf {\bibinfo
  {volume} {96}},\ \bibinfo {pages} {020406} (\bibinfo {year}
  {2017})}\BibitemShut {NoStop}%
\bibitem [{\citenamefont {Chiaracane}\ \emph {et~al.}(2021)\citenamefont
  {Chiaracane}, \citenamefont {Pietracaprina}, \citenamefont {Purkayastha},\
  and\ \citenamefont {Goold}}]{Chiaracane:2021}%
  \BibitemOpen
  \bibfield  {author} {\bibinfo {author} {\bibfnamefont {C.}~\bibnamefont
  {Chiaracane}}, \bibinfo {author} {\bibfnamefont {F.}~\bibnamefont
  {Pietracaprina}}, \bibinfo {author} {\bibfnamefont {A.}~\bibnamefont
  {Purkayastha}}, \ and\ \bibinfo {author} {\bibfnamefont {J.}~\bibnamefont
  {Goold}},\ }\href {\doibase 10.1103/PhysRevB.103.184205} {\bibfield
  {journal} {\bibinfo  {journal} {Physical Review B}\ }\textbf {\bibinfo
  {volume} {103}},\ \bibinfo {pages} {184205} (\bibinfo {year}
  {2021})}\BibitemShut {NoStop}%
\bibitem [{\citenamefont {Richter}\ \emph {et~al.}(2018)\citenamefont
  {Richter}, \citenamefont {Jin}, \citenamefont {De~Raedt}, \citenamefont
  {Michielsen}, \citenamefont {Gemmer},\ and\ \citenamefont
  {Steinigeweg}}]{Richter:2018}%
  \BibitemOpen
  \bibfield  {author} {\bibinfo {author} {\bibfnamefont {J.}~\bibnamefont
  {Richter}}, \bibinfo {author} {\bibfnamefont {F.}~\bibnamefont {Jin}},
  \bibinfo {author} {\bibfnamefont {H.}~\bibnamefont {De~Raedt}}, \bibinfo
  {author} {\bibfnamefont {K.}~\bibnamefont {Michielsen}}, \bibinfo {author}
  {\bibfnamefont {J.}~\bibnamefont {Gemmer}}, \ and\ \bibinfo {author}
  {\bibfnamefont {R.}~\bibnamefont {Steinigeweg}},\ }\href {\doibase
  10.1103/PhysRevB.97.174430} {\bibfield  {journal} {\bibinfo  {journal}
  {Physical Review B}\ }\textbf {\bibinfo {volume} {97}},\ \bibinfo {pages}
  {174430} (\bibinfo {year} {2018})}\BibitemShut {NoStop}%
\bibitem [{\citenamefont {Aleiner}\ \emph {et~al.}(2016)\citenamefont
  {Aleiner}, \citenamefont {Faoro},\ and\ \citenamefont {Ioffe}}]{Aleiner16}%
  \BibitemOpen
  \bibfield  {author} {\bibinfo {author} {\bibfnamefont {I.~L.}\ \bibnamefont
  {Aleiner}}, \bibinfo {author} {\bibfnamefont {L.}~\bibnamefont {Faoro}}, \
  and\ \bibinfo {author} {\bibfnamefont {L.~B.}\ \bibnamefont {Ioffe}},\ }\href
  {\doibase https://doi.org/10.1016/j.aop.2016.09.006} {\bibfield  {journal}
  {\bibinfo  {journal} {Annals of Physics}\ }\textbf {\bibinfo {volume}
  {375}},\ \bibinfo {pages} {378} (\bibinfo {year} {2016})}\BibitemShut
  {NoStop}%
\bibitem [{\citenamefont {Dymarsky}(2018)}]{Dymarsky2018}%
  \BibitemOpen
  \bibfield  {author} {\bibinfo {author} {\bibfnamefont {A.}~\bibnamefont
  {Dymarsky}},\ }\href {https://arxiv.org/abs/1804.08626v1} {\bibfield
  {journal} {\bibinfo  {journal} {arXiv:1804.08626 [cond-mat.stat-mech]}\ }
  (\bibinfo {year} {2018})}\BibitemShut {NoStop}%
\bibitem [{\citenamefont {Jansen}\ \emph {et~al.}(2019)\citenamefont {Jansen},
  \citenamefont {Stolpp}, \citenamefont {Vidmar},\ and\ \citenamefont
  {Heidrich-Meisner}}]{Vidmar:2019}%
  \BibitemOpen
  \bibfield  {author} {\bibinfo {author} {\bibfnamefont {D.}~\bibnamefont
  {Jansen}}, \bibinfo {author} {\bibfnamefont {J.}~\bibnamefont {Stolpp}},
  \bibinfo {author} {\bibfnamefont {L.}~\bibnamefont {Vidmar}}, \ and\ \bibinfo
  {author} {\bibfnamefont {F.}~\bibnamefont {Heidrich-Meisner}},\ }\href
  {\doibase 10.1103/PhysRevB.99.155130} {\bibfield  {journal} {\bibinfo
  {journal} {Physical Review B}\ }\textbf {\bibinfo {volume} {99}},\ \bibinfo
  {pages} {155130} (\bibinfo {year} {2019})}\BibitemShut {NoStop}%
\bibitem [{\citenamefont {Mierzejewski}\ and\ \citenamefont
  {Vidmar}(2020)}]{Vidmar2:2020}%
  \BibitemOpen
  \bibfield  {author} {\bibinfo {author} {\bibfnamefont {M.}~\bibnamefont
  {Mierzejewski}}\ and\ \bibinfo {author} {\bibfnamefont {L.}~\bibnamefont
  {Vidmar}},\ }\href {\doibase 10.1103/PhysRevLett.124.040603} {\bibfield
  {journal} {\bibinfo  {journal} {Physical Review Letters}\ }\textbf {\bibinfo
  {volume} {124}},\ \bibinfo {pages} {040603} (\bibinfo {year}
  {2020})}\BibitemShut {NoStop}%
\bibitem [{\citenamefont {Sch\"onle}\ \emph {et~al.}(2021)\citenamefont
  {Sch\"onle}, \citenamefont {Jansen}, \citenamefont {Heidrich-Meisner},\ and\
  \citenamefont {Vidmar}}]{Vidmar:2020}%
  \BibitemOpen
  \bibfield  {author} {\bibinfo {author} {\bibfnamefont {C.}~\bibnamefont
  {Sch\"onle}}, \bibinfo {author} {\bibfnamefont {D.}~\bibnamefont {Jansen}},
  \bibinfo {author} {\bibfnamefont {F.}~\bibnamefont {Heidrich-Meisner}}, \
  and\ \bibinfo {author} {\bibfnamefont {L.}~\bibnamefont {Vidmar}},\ }\href
  {\doibase 10.1103/PhysRevB.103.235137} {\bibfield  {journal} {\bibinfo
  {journal} {Physical Review B}\ }\textbf {\bibinfo {volume} {103}},\ \bibinfo
  {pages} {235137} (\bibinfo {year} {2021})}\BibitemShut {NoStop}%
\bibitem [{\citenamefont {Pappalardi}\ \emph {et~al.}(2020)\citenamefont
  {Pappalardi}, \citenamefont {Polkovnikov},\ and\ \citenamefont
  {Silva}}]{Pappalardi2020Quantum}%
  \BibitemOpen
  \bibfield  {author} {\bibinfo {author} {\bibfnamefont {S.}~\bibnamefont
  {Pappalardi}}, \bibinfo {author} {\bibfnamefont {A.}~\bibnamefont
  {Polkovnikov}}, \ and\ \bibinfo {author} {\bibfnamefont {A.}~\bibnamefont
  {Silva}},\ }\href {\doibase 10.21468/SciPostPhys.9.2.021} {\bibfield
  {journal} {\bibinfo  {journal} {SciPost Physics}\ }\textbf {\bibinfo {volume}
  {9}},\ \bibinfo {pages} {21} (\bibinfo {year} {2020})}\BibitemShut {NoStop}%
\bibitem [{\citenamefont {Sidje}(1998)}]{expokit}%
  \BibitemOpen
  \bibfield  {author} {\bibinfo {author} {\bibfnamefont {R.~B.}\ \bibnamefont
  {Sidje}},\ }\href {https://www.maths.uq.edu.au/expokit/paper.pdf} {\bibfield
  {journal} {\bibinfo  {journal} {ACM Transactions on Mathematical Software}\
  }\textbf {\bibinfo {volume} {24}},\ \bibinfo {pages} {130} (\bibinfo {year}
  {1998})}\BibitemShut {NoStop}%
\bibitem [{\citenamefont {Brenes}\ \emph {et~al.}(2019)\citenamefont {Brenes},
  \citenamefont {Varma}, \citenamefont {Scardicchio},\ and\ \citenamefont
  {Girotto}}]{brenes2019massively}%
  \BibitemOpen
  \bibfield  {author} {\bibinfo {author} {\bibfnamefont {M.}~\bibnamefont
  {Brenes}}, \bibinfo {author} {\bibfnamefont {V.~K.}\ \bibnamefont {Varma}},
  \bibinfo {author} {\bibfnamefont {A.}~\bibnamefont {Scardicchio}}, \ and\
  \bibinfo {author} {\bibfnamefont {I.}~\bibnamefont {Girotto}},\ }\href
  {\doibase https://doi.org/10.1016/j.cpc.2018.08.010} {\bibfield  {journal}
  {\bibinfo  {journal} {Computer Physics Communications}\ }\textbf {\bibinfo
  {volume} {235}},\ \bibinfo {pages} {477} (\bibinfo {year}
  {2019})}\BibitemShut {NoStop}%
\bibitem [{\citenamefont {Hochbruck}\ and\ \citenamefont
  {Lubich}(1997)}]{expokiterror}%
  \BibitemOpen
  \bibfield  {author} {\bibinfo {author} {\bibfnamefont {M.}~\bibnamefont
  {Hochbruck}}\ and\ \bibinfo {author} {\bibfnamefont {C.}~\bibnamefont
  {Lubich}},\ }\href {https://na.math.kit.edu/download/papers/exp.pdf}
  {\bibfield  {journal} {\bibinfo  {journal} {SIAM Journal on Numerical
  Analysis}\ }\textbf {\bibinfo {volume} {34}},\ \bibinfo {pages} {1991}
  (\bibinfo {year} {1997})}\BibitemShut {NoStop}%
\end{thebibliography}%

\end{document}